\documentclass[prc,twocolumn,amsmath,amssymb,superscriptaddress,nofootinbib,preprintnumbers]{revtex4-1}
\usepackage{url,color,colordvi,epsfig,pstricks}
\usepackage[colorlinks=true, citecolor=blue, filecolor=blue, linkcolor=blue, urlcolor=blue, pdftex]{hyperref}
\usepackage{graphicx,bm,dcolumn}
\usepackage{hyperref,slashed,lineno}
\usepackage{physics}
\usepackage{inputenc}

\usepackage{bbold} 

\hypersetup{
    colorlinks=true, 
    linktoc=all, 
    linkcolor=blue, 
    citecolor=blue}

\newcommand{\ba}{\begin{eqnarray}}
\newcommand{\ea}{\end{eqnarray}}

\newcommand{\nk}{{\bf      k}}
\newcommand{\np}{{\bf      p}}
\newcommand{\nq}{{\bf      q}}
\newcommand{\nr}{{\bf      r}}

\newcommand{\nt}{{\bf      t}}

\newcommand{\nz}{{\bf      z}}

\begin{document}

\preprint{NT@UW-25-20}

\title{Factorized distorted wave calculations for electron, neutrino and BSM processes}
\author{A.~Nikolakopoulos}\email{anikolak@uw.gov}
\affiliation{Physics Department, University of Washington, Seattle WA}
\author{R.~Gonz\'alez-Jim\'enez}\email{raugj@us.es}
\affiliation{Departamento de Física Atómica, Molecular y Nuclear, Universidad de Sevilla, 41080 Sevilla, Spain}

\begin{abstract}
We point out that, under certain conditions, the nuclear currents that couple to vector bosons can be written as the trace of the product of two matrices.
One contains `nucleon dynamics', e.g. form factors, the other contains the overlaps of nuclear wavefunctions.
This factorized form may always be obtained and viewed as a `local' approximation, in which particle four-momenta that enter in the transition operator are fixed to their asymptotic values.
We write the overlap matrix in a general form in terms of Dirac matrices. The current with arbitrary couplings to the nucleon can then be evaluated using standard trace identities.
We show that this encompasses non-relativistic models as well.
We have tabulated overlaps obtained in the relativistic distorted-wave impulse approximation, which can be used to compute the single-nucleon knockout cross section. We give a self-contained overview of the formulae. 
We discuss some properties of the overlap matrices which may be derived from general principles, and highlight differences with the commonly used plane-wave impulse approximation.
The factorized form is attractive for (neutrino) event generators: it abstracts away the nuclear model and allows to easily modify couplings to the nucleon. 
This allows to consistently treat electron, neutrino and beyond Standard Model (BSM) processes.
\end{abstract}

\maketitle

\section{Introduction}
Accelerator based neutrino experiments require flexible  realistic descriptions of electroweak interactions with the nucleus~\cite{NUSTECWP, KatoriMartinireview}.
A simple way to achieve a flexible description is to adopt the plane-wave impulse approximation (PWIA), in which the cross section can be written as the product of a single-nucleon cross section with the diagonal hole-spectral function of the nucleus~\cite{Boffi93}.
As such, different electroweak and beyond Standard Model (BSM) processes involving a single nucleon, e.g. nucleon knockout and single-pion production, can be described by changing the single-nucleon cross section.
This means that it is straightforward to, for example, modify the nucleon vector and axial form factors and provide consistently cross sections for electron and neutrino interactions.
The issue with the PWIA is  that the final-state nucleon is treated as a plane wave; final-state interactions (FSI) are completely neglected.
The PWIA therefore fails on several points. For example, the position of the quasielastic peak is shifted to too large energy transfer, Pauli-blocking is not naturally included, and there are ambiguities on how to account for binding energy and vector-current conservation.

These issues are due to the fact that the initial and final states in matrix elements computed in the PWIA are eigenstates of a different Hamiltonian.
They can be partly resolved by adopting the (relativistic) distorted wave impulse approximation, (R)DWIA.
In this case, initial and final states could be described as energy eigenstates of the same strong potential.
This results in a shift of the quasielastic peak compared to the PWIA because of the potential felt by the final-state nucleon.
It furthermore guarantees that the initial and final-states are orthogonal, which means that Pauli's exclusion principle is naturally incorporated~\cite{Nikolakopoulos:2019qcr, Gonzalez-Jimenez:2019qhq}.  
Finally, the Dirac current is conserved in this case, $\partial_\mu \overline{\psi} \gamma^\mu \psi = 0$, which allows to construct a conserved vector current~\cite{SEROT1981263, Nikolakopoulos:2020fti, Serot:1997xg} that naturally incorporates binding if the small nuclear recoil is neglected.

After a number of theoretical works applying the theory for neutrino induced single-nucleon knockout~\cite{Gonzalez-Jimenez13c, Gonzalez-Jimenez:2021ohu, Franco-Patino:2022tvv, Nikolakopoulos:2022qkq, Nikolakopoulos:2024mjj, PhysRevC.109.045502}, the NEUT event generator has adopted the RDWIA approach~\cite{McKean:2025khb}.
The implementation is done by supplying tables of hadron tensor elements.
While this is an improvement, this approach is not as flexible as the PWIA.
For example, the hadron tensor is computed with fixed vector and axial form factors.
These form factors cannot easily be modified.

Here we point out that, under suitable conditions, one may obtain factorization of the hadron current instead of the hadron tensor.
We show this for the RDWIA, where the factorized current may be written as the trace of a product of two $4\times 4$ matrices. 
One contains the `single-nucleon dynamics' (e.g. form factors) and depends on the type of interaction. The other contains overlap integrals of single-nucleon wave functions.
The factorized form adopted here is quite general and does not rely specifically on the RDWIA.
We show in particular that typical non-relativistic mean-field approaches may be factorized in exactly the same manner.

Currents that are factorizable in this way have been used in neutrino-induced single-nucleon knockout. 
One may always obtain factorization by adopting the `local' or `asymptotic approximation', as done for example for single-pion production in Refs.~\cite{Garcia-Marcos:2023rnj, Nikolakopoulos:2022tut}.

An implementation of the formalism in neutrino event generators offers the flexibility of the PWIA: a formulation of the amplitude where the nuclear overlaps are separated from the process considered. 
The overlap matrices may be tabulated to allow for fast RDWIA calculations in neutrino and electron interactions where the nucleon form factors can be modified at will.
Other processes in which arbitrary couplings to the nucleon are considered can be readily included.
Since the same factorized form may be used for many different theoretical approaches to the nuclear overlaps, one does not need to consider a specific nuclear model for a general implementation.

In this work we consider single-nucleon knockout. 
We derive some properties of the nuclear overlap matrix and discuss their implications for the cross section.
We have computed and tabulated the nuclear overlaps for carbon and oxygen obtained in the RDWIA using the Energy-Dependent Relativistic Mean Field (EDRMF) potential of Refs.~\cite{Gonzalez-Jimenez:2019qhq, Gonzalez-Jimenez:2019ejf}.
These may be obtained from~\cite{Factorized_RDWIA_Code}, along with a code that computes the cross section for electron and charged-current neutrino interactions.
The extension of the formalism to include the interference with two-body currents~\cite{Franco-Munoz:2022jcl, Franco-Munoz:2023zoa} and processes where additional particles are produced in the hadron system, e.g. single-pion production, will be pursued in future work.

This paper is structured as follows.
In section~\ref{sec:Factorized} we discuss the conditions for factorization of the hadron current, and introduce the overlap matrices. 
Section~\ref{sec:Crosssection} is a self-contained overview of the calculation of electron and neutrino-induced single-nucleon knockout in terms of overlap matrices.
Then, in section~\ref{sec:properties}, we discuss some general properties of the overlap matrix and their implications for the cross section.
In section~\ref{sec:nonrel} in particular, we show how factorization is obtained in non-relativistic approaches.
Conclusions and outlook are given in section~\ref{sec:conclusions}.

\section{Factorized RDWIA calculation}
\label{sec:Factorized}
The invariant matrix elements for semi-leptonic processes where a single boson is exchanged with the target nucleus, can be written as a product of hadron and lepton currents, $J^{\mu}_l$ and $J^{\mu}_h$,
\begin{equation}
\mathcal{M}_{\lambda,\lambda^\prime} = J_{l}(\lambda) \cdot J_h(\lambda^\prime).
\end{equation}
Here $\lambda, \lambda^\prime$ denote different polarizations of the lepton and hadron systems. 
The squared matrix element summed over polarizations can then be written in terms of the lepton and hadron tensors
\begin{equation}
\sum_{\lambda,\lambda^\prime} \rvert \mathcal{M}_{\lambda,\lambda^\prime} \lvert^2 = L_{\mu\nu} H^{\mu\nu},
\end{equation}
defined as
\begin{equation}
H^{\mu\nu} = \sum_{\lambda} \left[J_h^\mu(\lambda) \right]^* J_h^{\nu}(\lambda),
\end{equation}
and equivalently for the lepton tensor $L_{\mu\nu}$.

If one adopts the plane wave impulse approximation (PWIA), the hadron tensor factorizes and is straightforward to compute.
For example, when a single nucleon with mass $M_N$ is knocked out from a nucleus with mass $M_A$, thereby leaving a residual system with mass $M_B = M_A - M_N + E_m$ and momentum $\mathbf{k}_B$ the hadron tensor in the PWIA factorizes as~\cite{Boffi93}
\begin{equation}
H^{\mu\nu}_{PWIA} = h_{s.n.}^{\mu\nu} \times S(E_m, \lvert \mathbf{p}_m\rvert),
\end{equation}
where $\mathbf{p}_m \equiv -\mathbf{k}_B$ is the missing momentum, and $E_m$ the missing energy.
Here $h_{s.n.}^{\mu\nu}$ is the single-nucleon hadron tensor, which can be computed analytically~\cite{VanOrden:2019krz, Nikolakopoulos:2024mjj,PhysRevC.102.064626}. 
All nuclear effects are contained in the diagonal single-nucleon spectral function $S(E_m, |\np_m|)$~\cite{dickhoff2008many, Boffi93}, which depends only on the momentum and energy of the residual system.
This factorization is attractive, since one can easily modify $h^{\mu\nu}_{s.n.}$ to treat different one-nucleon interactions, see e.g.~\cite{Rocco:2019gfb, Kopp:2024yvh}. 

In general, this factorization of the hadron tensor does not occur.
Under suitable conditions however, one can obtain factorization of the hadron current instead.  
We assume that the hadron current, in the impulse approximation where momentum $\mathbf{t}$ is transferred to a single nucleon, can be written as the integral over a bilinear of single-particle wavefunctions
\begin{equation}
\label{eq:Current_bilinear}
J^{\mu}(\kappa,s,m_j) = \int \mathrm{d} \mathbf{r} e^{i\mathbf{r}\cdot\mathbf{t}} \overline{\Psi}^s(\mathbf{r}, \mathbf{k}_N) \Gamma^{\mu}(\mathbf{r}) \phi_{\kappa}^{m_j}(\mathbf{r}),
\end{equation}
where we suppress the dependence of $J^{\mu}$ on momenta\footnote{Momenta other than $\mathbf{t}$ and $\mathbf{k}_N$ may enter in the current. 
For example, in incoherent single-pion production $\Gamma^{\mu}$ also depends on the four-momentum of the pion $K_\pi$, and $\nt=\nq-\nk_\pi$ with $\nq$ being the momentum of the virtual boson~\cite{Garcia-Marcos:2023rnj}.}.
Here $\overline{\Psi}^s(\mathbf{r},\mathbf{k}_N)$ is a scattering state for the knocked out nucleon with asymptotic momentum $\mathbf{k}_N$ and spin projection $s$. 
The residual system is described by the hole state $\phi_\kappa^{m_j}$, with good parity and total angular momentum $j$. 
Here, $\kappa$ labels the total angular momentum $j = \lvert \kappa \rvert - 1/2$, and orbital angular momentum $l=j\pm 1/2$ for $\kappa = \pm \lvert \kappa \rvert$; 
$m_j$ is the projection of $j$.
This is the form of the hadron current in e.g. the RDWIA, where the states can be represented by four-component spinors, and $\Gamma^\mu$ as a $4\times 4$ matrix.
We use this explicit representation in the following, but one does not have to assume any specific model.
The only assumption is that the residual system is left in a state with good total angular momentum and that there exists a scattering state, both of which can be represented by four-component spinor wavefunctions.

Then, if the bilinear operator $\Gamma^{\mu}$ does not depend on $\mathbf{r}$ 
we obtain a factorized hadron current 
\begin{equation}
\label{eq:Factorized}
J^{\mu}(\kappa,s,m_j) = \mathrm{Tr}\left[\Gamma^{\mu} S(\mathbf{t},\mathbf{k}_N ; \kappa,s,m_j) \right],
\end{equation}
where
\begin{align}
\label{eq:def_Skappa}
S(\mathbf{t},\mathbf{k}_N ; \kappa,s,m_j) =  \int \mathrm{d} \mathbf{r} e^{i\mathbf{r}\cdot\mathbf{t}} \phi_{\kappa}^{m_j}(\mathbf{r}) \overline{\Psi}^s(\mathbf{r},\nk_N).
\end{align}

The assumption of the factorization in Eq.~(\ref{eq:Factorized}) is of course the assumption of a model. Specifically, it is an assumption on how the amplitudes for scattering of free nucleons are introduced in the nuclear medium~\cite{Caballero:1997gc, SEROT1981263, Udias93}.
Forms that are factorizable in this way have been empirically successful, e.g. in electromagnetic single-nucleon knockout~\cite{Udias93}. 
In this case one defines
\begin{equation}
\Gamma^\mu = F_1(Q^2) \gamma^\mu - \frac{F_2(Q^2)}{4M_N} Q_\nu \left[\gamma^\mu, \gamma^\nu \right],
\end{equation} 
where $Q^\mu = (\omega, \mathbf{q})$ is taken to be the four-momentum of the exchanged photon, and $\mathbf{t} = \mathbf{q}$.
This form has been used also for neutrino interactions with the addition of the axial current using Eq.~(\ref{eq:Gamma_axial}).

This factorization holds for typical non-relativistic approaches as well.
In that case, one performs an expansion of $\overline{u} \Gamma^\mu u$, where $u$ are free positive-energy Dirac spinors~\cite{Jeschonnek:1997dm, walecka2012semileptonic, Waleckapaper, Udias:95, Amaro07}. 
This yields a momentum dependent operator, which may be evaluated with non-relativistic wavefunctions. 
As illustrated in Sec.~\ref{sec:nonrel}, this procedure may be mapped to overlap matrices $S(\kappa,s,m_j)$ constructed from positive energy four-component spinors with a momentum-independent $\Gamma^\mu$.
One then obtains the same factorization property as in Eq.~(\ref{eq:Factorized}).

In the general case, in which $\Gamma^\mu$ depends on $\nr$, one may recover Eq.~(\ref{eq:Factorized}) in the `asymptotic approximation', also called `local approximation', where the $r$-dependent momenta in $\Gamma^\mu$ are replaced by their fixed  asymptotic values.
See, for example, Refs.~\cite{Garcia-Marcos:2023rnj, Nikolakopoulos:2022tut, LAGET197281, VanderhaeghenPhD, PhysRevC.48.816, PhysRevC.28.1725} for  discussions of this approximation in the context of coherent and incoherent single-pion production.
In any case, it is possible to extend the factorization to momentum-dependent operators by introducing derivatives in Eq.~(\ref{eq:def_Skappa}), but this is beyond the scope of the present work.

The overlap matrix may be decomposed into any basis, a suitable decomposition is into scalar, pseudo-scalar, vector, pseudo-vector, and tensor components:
\begin{align}
\label{eq:Fierzlike}
&S(\mathbf{t},\mathbf{k}_N ; \kappa,s,m_j) = \nonumber \\
&\frac{1}{4}\left(s \mathbb{1} + p \gamma^5 + V^\mu \gamma^\mu + A^\mu \gamma^\mu\gamma^5 + T^{\mu\nu} \frac{1}{2}\left[ \gamma^\mu, \gamma^\nu \right]\right),
\end{align}
where summation over repeated indices is implied and, to shorten the notation, we suppress the dependence on (angular) momenta of the expansion coefficients $s, p, V^\mu, A^\mu$ and $T^{\mu\nu}$.
The trace of Eq.~(\ref{eq:Factorized}) can then be performed analytically using standard trace identities.
This provides the same flexibility as in the PWIA, i.e. one can evaluate the current with arbitrary couplings to the nucleon.

In-situ evaluation of the overlap matrix $S$ is often too costly to efficiently generate scattering events, which may limit the usability in (neutrino) event generators.
The overlaps can be tabulated to allow for rapid evaluation.
In principle, $S(\mathbf{t},\mathbf{k}_N ; \kappa, s, m_j)$ is a function of the momentum transfer, the nucleon momentum and the quantization axis.
The latter is of course arbitrary, since we anticipate summing over all angular momenta. 
We may align it with any of the vectors, e.g. the nucleon momentum, such that $S$ only depends on $\mathbf{t}$ and $\lvert \mathbf{k}_N\rvert$.
Furthermore, for single-nucleon knockout in the impulse approximation, the momentum transferred to the nucleon $\mathbf{t}$ is the total momentum transferred to the nuclear system. 
In this case, only the dependence on $\lvert \mathbf{t} \rvert$, $\lvert \mathbf{k}_N \rvert$ and the relative angle $\hat{\mathbf{t}}\cdot\hat{\mathbf{k}}_N$ is required.
We have computed and tabulated the components of Eq.~(\ref{eq:Fierzlike}) obtained using the EDRMF potential of Refs.~\cite{Gonzalez-Jimenez19, Gonzalez-Jimenez:2021ohu}. 
They can be obtained from~\cite{Factorized_RDWIA_Code}.
A version of the code developed in \cite{UdiasPhD} was used to obtain the nucleon scattering state.
A self-contained overview of the calculation of the single-nucleon knockout cross section in terms of these tables is given in the next section.

\section{Neutrino and electron induced single-nucleon knockout}
\label{sec:Crosssection}
\begin{figure}
    \centering
    \includegraphics[width=0.45\textwidth]{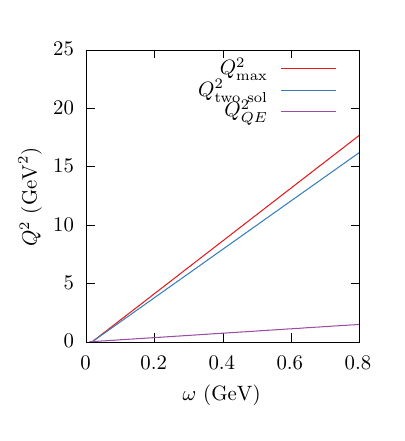}
    \caption{Phase-space boundaries from Eq.~(\ref{eq:q2_bound_twosol}). Above $Q^2_{\mathrm{max}}(\omega)$ there are no physical solutions $\mathbf{k}_N$ to Eq.~(\ref{eq:Econservation}). For $Q^2 < Q^{2}_{\mathrm{two~sol}}(\omega)$ at most one physical solution is possible. Above this bound, two physical solutions are possible. The approximate position of the quasielastic peak is $Q_{QE}^2 = 2M_N\omega$. We used $E_\kappa = 16~\mathrm{MeV}$. }
    \label{fig:phase_space}
\end{figure}
We describe the computation of the cross section for single-nucleon knockout from a shell labeled by $\kappa$ and with missing energy $E_\kappa$, in terms of the overlap matrix of Eq.~(\ref{eq:def_Skappa}).
The process considered is
\begin{equation}
l(k) + A (k_A) \rightarrow l^\prime(k^\prime) + N(k_N) + B(k_B),
\end{equation}
here $A$ is the initial nucleus, and $B$ the final-state nucleus with mass $M_B = M_A + E_\kappa - M_N$.
The four-momentum transfer to the nucleus is $Q = k - k^\prime = (\omega , \mathbf{q})$, where $k$, $k^\prime$ are the initial and final lepton four-momenta.
We define $Q^2 \equiv -Q\cdot Q$ to be positive.
In the laboratory frame, where the target nucleus is at rest, the energy of the knocked out nucleon is determined by
\begin{equation}
\label{eq:Econservation}
\sqrt{\mathbf{k}_N^2 + M_N^2}  = M_N + \omega - E_\kappa - \sqrt{\mathbf{p}_m^2 + M_B^2} - M_B,
\end{equation}
where the missing momentum $\mathbf{p}_m \equiv -\mathbf{k}_B= \mathbf{k}_N - \mathbf{q}$.
The cross section can be written as
\begin{equation}
\label{eq:sigma_diff}
\frac{\mathrm{d}\sigma(E)}{\mathrm{d}E^\prime \mathrm{d}\Omega^\prime \mathrm{d}\Omega_N} = \mathcal{F}^2_X \frac{\lvert \mathbf{k}^\prime\rvert}{E} \frac{\lvert \mathbf{k}_N \rvert E_N}{(2\pi)^5f_{rec}} L_{\mu\nu} H^{\mu\nu},
\end{equation}
where the recoil factor is
\begin{equation}
f_{rec} = \big\lvert 1 + \frac{E_N}{M_B} \left(1 - \frac{\mathbf{q}\cdot\mathbf{k}_N}{\mathbf{k}_N^2} \right) \big\rvert .
\end{equation}
In some cases, Eq.~(\ref{eq:Econservation}) allows for two solutions for the nucleon momentum given a nucleon scattering angle and four-momentum transfer $Q$.
As shown in appendix~\ref{app:kinematics}, this is possible only in a narrow kinematic region around the edge of the allowed phase space where
\begin{equation}
\label{eq:q2_bound_twosol}
    (\omega + M_A - M_N)^2 - M_B^2< \mathbf{q}^2 < (\omega + M_A)^2 - (M_N+M_B)^2.
\end{equation}
For $\lvert \mathbf{q} \rvert$ larger than this upper bound, there are no solutions. 
For fixed $Q^2$, these bounds correspond to 
\begin{equation}
  \frac{Q^2}{2M_A}  < \omega - E_\kappa \lesssim \frac{Q^2}{2M_B}, 
\end{equation} 
where the upper bound is approximate, see Eq.~(\ref{eq:w_approx_Q2MB}).
This kinematic region corresponds to the case in which the nucleon kinetic energy is comparable to that of the remnant system in elastic scattering with four-momentum transfer $Q^2$.
Both solutions can be taken into account in this kinematic region, or one may choose to reject the solution where the residual system absorbs a large amount of momentum.
The phase space is shown in Fig.~\ref{fig:phase_space}, the contribution from this kinematic region is small, since it lies far from the quasielastic peak.

The lepton tensor is defined as
\begin{align}
L^{\mu\nu} &= k^\mu k^{\prime \nu} + k^{\prime \mu} k^{\nu} +  g^{\mu\nu}(mm^\prime - k^\prime\cdot k) \nonumber \\
&+ ih \epsilon^{\mu\nu\alpha\beta } k^\prime_{\alpha} k_{\beta}.
\label{eq:Leptontensor}
\end{align}
where for electromagnetic interactions $h=0$, i.e. we assume unpolarized electrons, and the sum over polarizations leads to the disappearance of the antisymmetric contribution to the lepton tensor. 
For neutrino and antineutrino interactions $h=-1$ and $h=1$, respectively.
The prefactors in Eq.~(\ref{eq:sigma_diff}) are 
\begin{align}
\mathcal{F}^2_{EM} = \frac{1}{2}\left(\frac{4\pi\alpha}{Q^2}\right)^2, \quad \mathcal{F}^2_{CC} = G_F^2\cos^2\theta_c,
\end{align}
for electromagnetic and charged-current interactions respectively. 
Here $\alpha$ is the fine-structure constant, $G_F$ the Fermi coupling constant, and $\cos\theta_c$ the Cabibbo angle.
The hadron tensor is defined as
\begin{align}
&H^{\mu\nu}(Q,\theta_N, \phi_N) \\
&= \sum_{s_N,m_j} \left[J^{\mu}(Q,\theta_N,\phi_N; s_N, m_j) \right]^* J^{\nu}(Q, \theta_N, \phi_N; s_N, m_j).\nonumber
\end{align}
The dependence of the hadron tensor on the nucleon azimuth angle $\phi_N$, i.e. the angle between the lepton plane and hadron plane, can be obtained by a rotation of the hadron tensor around $\mathbf{q}$ with the angle $\phi_N$. The $\phi_N$ dependence of the cross section can then be written in terms of sines and cosines. 
If we choose $\mathbf{q}$ to determine the $z$-axis and denote the symmetric and anti-symmetric parts of the hadron tensor as
\begin{equation}
\label{eq:W_0_sym}
W_{s}^{\mu\nu} \equiv \mathrm{Re}\left[ H^{\mu\nu}(Q,\theta_N, \phi_N=0)  \right]
\end{equation}
and
\begin{equation}
\label{eq:W_0_asym}
W_{a}^{\mu\nu} \equiv \mathrm{Im}\left[ H^{\mu\nu}(Q,\theta_N, \phi_N=0)  \right],
\end{equation}
and similarly for the lepton tensor
\begin{equation}
    L_{{\mu\nu}} = L_{\mu\nu}^s + iL_{\mu\nu}^a,
\end{equation}
the contraction of lepton and hadron tensors can be written as~\cite{Sobczyk:angles}
\begin{widetext}
\begin{align}
\label{eq:LH_incl}
L_{\mu\nu} H^{\mu\nu} &= L_{00}^{s} W^{00}_s + 2 L_{03}^{s} W^{03}_s + L_{33}^{s} W^{33}_s + \frac{L^s_{11} + L^s_{22}}{2} \left( W_{s}^{11} + W_{s}^{22} \right) - 2L_{12}^{a} W_{a}^{12}, \\
&+2\left[L_{01}^{s} W_s^{01} + L_{13}^{s} W_s^{13} - L_{02}^{a} W_a^{02} - L_{23}^{a} W_a^{23} \right]\cos\phi_N +\frac{\left(L_{11}^{s} - L_{22}^{s} \right)}{2}\left( W^{11}_s - W^{22}_s \right) \cos 2\phi_N, \label{eq:LH_cosine} \\
&-2\left[L_{01}^{s} W_s^{02} + L_{13}^{s} W_s^{23} + L_{02}^{a} W_a^{01} + L_{23}^{a} W_a^{13} \right]\sin\phi_N +\left(L_{22}^{s} - L_{11}^{s} \right)W_{s}^{12} \sin 2\phi_N.
\label{eq:LH_sine}
\end{align}
\end{widetext}
Note that we retain the terms $W_s^{02}, W_s^{23}, W_s^{12}$. They are found to be non-zero in our RDWIA calculations. This is discussed in appendix~\ref{app:funny_tensors}.

We now write the hadron-tensor elements in terms of the overlap matrix $S(\mathbf{q}, \mathbf{k}_N  ;\kappa, s_N, m_j)$. The latter are computed in a reference system where $\mathbf{k}_N$ determines the $z$-axis and $\nq$ lies in the $xz$-plane\footnote{As discussed in the next section, this choice yields a simpler form for the scattering wavefunction of the nucleon.}. 
The hadron tensor elements $W^{\mu\nu}$ are related to those obtained in that system, denoted $\tilde{W}^{\mu\nu}$, by a rotation of angle $\theta_N$ around the $y$-axis:
\begin{equation}
W^{\mu\nu} = \left[R^{\mu}_{~\alpha}(\theta_N)\right]^* R^{\nu}_{~\beta}(\theta_N) \tilde{W}^{\alpha,\beta}.
\end{equation} 
We denote the four-momentum transfer in this system as $\tilde{Q}^\mu = R^{\mu}_{~\alpha}(-\theta_N) Q^{\alpha}$.
We may now compute the hadron tensor
\begin{align}
\tilde{W}^{\mu\nu} = \sum_{s_N, m_j} \left[\tilde{J}^{\mu}(\tilde{Q}, \lvert \mathbf{k}_N \rvert, s_N, m_j)\right]^* \tilde{J}^{\nu}(\tilde{Q}, \lvert \mathbf{k}_N \rvert,s_N, m_j),
\end{align}
for neutrino and electron interactions.
We define
\begin{align}
\label{eq:Gamma_vector}
\Gamma^{\mu}(Q) &= F_1(Q^2) \gamma^\mu - \frac{F_{2}(Q^2)}{4M_N} \left[ \gamma^{\mu}, \gamma^{\nu} \right] Q_{\nu} \\
&- G_{A}(Q^2) \gamma^{\mu}\gamma^5 - \frac{G_{P}(Q^2)}{2M_N} Q^{\mu} \gamma^5
\label{eq:Gamma_axial}
\end{align}
in terms of four form factors\footnote{Note that this form does not guarantee a conserved axial current in the limit of vanishing pion mass in the presence of a nuclear potential. To construct an axial current with this property one could replace the pseudoscalar term $Q^\mu \gamma^5 G_P(Q^2)/(2M_N) \rightarrow  Q^{\mu} \slashed{Q}\gamma^5 G_A(Q^2)/(m_\pi^2 + Q^2)$ in Eq.~(\ref{eq:Gamma_axial}). For free nucleons these forms are equivalent since $\bar  u(\mathbf{p}+\mathbf{q}) \slashed{Q} \gamma^5 u(\mathbf{p}) = 2M_N \bar u(\mathbf{p} + \mathbf{q}) \gamma^5 u(\mathbf{p})$.}.

For electron interactions with the proton (neutron) $F_{i} = F_{i}^{p} (F_{i}^n)$, are the Dirac and Pauli form factors, and the axial form factors $G_{A} = G_{P} = 0$.
For charged current interactions the axial form factors do contribute and $F_{i} = F_{i}^V = F_{i}^p - F_{i}^n$ are the isovector vector form factors.
The extension to weak neutral currents is straightforward~\cite{Ivanov15,Gonzalez-Jimenez13} but not considered here.
The hadron current is then
\begin{equation}
\tilde{J}^{\mu}(s_N, m_j) = \mathrm{Tr} \left[ \Gamma^{\mu}(\tilde{Q}) S(s_N, m_j) \right],
\end{equation}
where we have suppressed the dependence on momenta and $\kappa$.
The overlap $S(s_N, m_j)$ is completely determined by (pseudo)scalar, (axial-)vector and tensor components of Eq.~(\ref{eq:Fierzlike}) given by the traces
\begin{equation}
\label{eq:scalar}
s_{s_N,m_j} \equiv \mathrm{Tr}\left[S(s_N,m_j)\right], 
\end{equation}
\begin{equation}
p_{s_N,m_j} \equiv \mathrm{Tr}\left[\gamma^5 S(s_N,m_j)\right], 
\end{equation}
\begin{equation}
V^{\mu}_{s_N,m_j} \equiv \mathrm{Tr}\left[\gamma^\mu S(s_N,m_j)\right], 
\end{equation}
\begin{equation}
A^{\mu}_{s_N,m_j} \equiv \mathrm{Tr}\left[\gamma^\mu\gamma^5 S(s_N,m_j)\right], 
\end{equation}
\begin{equation}
\label{eq:Tensor}
T^{\mu\nu}_{s_N,m_j} \equiv \mathrm{Tr}\left[\frac{1}{2}\left[\gamma^\mu,\gamma^\nu\right] S(s_N,m_j)\right]. 
\end{equation}
With Eqs.~(\ref{eq:Gamma_vector}-\ref{eq:Gamma_axial}) the current is given by
\begin{align}
\tilde{J}^\mu(s_N,m_j) &= F_1(Q^2) V^{\mu}_{s_N,m_j} - \frac{F_2(Q^2)}{2M_N} \tilde{Q}_{\nu} T^{\mu\nu}_{s_N,m_j} \nonumber \\
&- G_A(Q^2) A^{\mu}_{s_N,m_j} - \frac{G_{P}(Q^2)}{2M_N} \tilde{Q}^\mu p_{s_N,m_j}.
\label{eq:currentVTAP}
\end{align}

One can tabulate the coefficients of Eqs.~(\ref{eq:scalar}-\ref{eq:Tensor}) which may be used to produce the single-nucleon knockout cross section with arbitrary form factors.
Given our choice of reference system, these only depend on $\lvert \mathbf{k}_N \rvert$, $\lvert \mathbf{q} \rvert$ and the angle $\hat{\mathbf{k}}_N\cdot\hat{\mathbf{q}}$.
To tabulate them, it is better to use the equivalent set $\lvert \mathbf{p}_m\rvert, \lvert \mathbf{k}_N\rvert, \hat{\mathbf{p}}_m\cdot\hat{\mathbf{k}}_N$ instead. 
This is more convenient since the overlaps become negligible for large missing momenta $\lvert \mathbf{p}_m \rvert \gtrsim 600~\mathrm{MeV}$ and the angle $\hat{\mathbf{p}}_m\cdot\hat{\mathbf{k}}_N$ provides better coverage of the physical region than $\hat{\mathbf{q}}\cdot\hat{\mathbf{k}}_N$.

Tables obtained using the EDRMF potential, for single-nucleon knockout from carbon and oxygen may be obtained from~\cite{Factorized_RDWIA_Code} along with a simple code that computes the cross section.
We provide tables for proton and neutron final-states for neutral current interactions, i.e.,  $p\rightarrow p$ and $n\rightarrow n$; and charged-current interactions, i.e., $p\rightarrow n$ and $n\rightarrow p$. We point out that the Coulomb potential is included when the scattering state is a proton.
This code is equivalent to the mean-field contribution of Ref.~\cite{McKean:2025khb}, with the advantage that, in the present approach, arbitrary nucleon form factors can be considered.
Moreover, this code can be adapted to include arbitrary forms of $\Gamma^\mu$, thereby facilitating studies of BSM interactions. 

\section{Properties of the overlap matrix $S$ and approximations}
\label{sec:properties}
We discuss some of the properties of the overlap matrix $S$, first assuming the case where initial and final-state wavefunctions are written in terms of relativistic angular-momentum states.
This is particularly the case for the RDWIA calculations provided in this work, but the relations are quite general and one does not need to assume any specific model.

We use following form for angular momentum states~\cite{Greiner}
\begin{equation}
\label{eq:psi_radial}
\psi_{\kappa}^{m}(\mathbf{r}) =
\begin{pmatrix}
g_\kappa(r) \Phi_{\kappa}^{m}\left(\Omega_r \right) \\
if_\kappa(r) \Phi_{-\kappa}^m\left( \Omega_r \right)
\end{pmatrix}.
\end{equation}
Here, the total angular momentum $j = \lvert \kappa \rvert - 1/2$, and orbital angular momentum $l= j \pm 1/2$ for $\kappa = \pm \lvert \kappa \rvert$.
The spin-spherical harmonics are defined in terms of Clebsch-Gordan coefficients and spherical harmonics as
\begin{equation}
\label{eq:SSH}
\Phi_\kappa^{m}(\Omega_r) = \sum_{m_l, m_s} \langle l, m_l; 1/2, m_s \vert j, m \rangle Y_{l,m_l}(\Omega_r) \chi^{m_s},
\end{equation}
with $\chi^{\pm1/2}$ orthonormal spin states.
A scattering state can be written in terms of angular momentum states by using a partial-wave expansion 
\begin{align}
\label{eq:psi_scatter_pwe}
&\Psi^{s_N}(\mathbf{r}, \mathbf{k}_N) = \nonumber \\
&\sum_{\kappa,m_l,m_j} e^{-i\delta_\kappa} i^{l} \langle l, m_l; 1/2, s_N \vert j, m \rangle Y_{l,m_l}(\Omega_N) \psi^{m_j}_\kappa(\mathbf{r}, \lvert \mathbf{k}_N \rvert),
\end{align}
where the dependence on $\lvert \mathbf{k}_N \rvert$ enters in the radial wavefunctions $g_\kappa(r,\lvert \mathbf{k}_N \rvert)$, $f_\kappa(r, \lvert \mathbf{k}_N \rvert)$ and phase shifts $\delta_\kappa(\lvert \mathbf{k}_N \rvert)$.
We then write
\begin{equation}
S(\mathbf{q}, \mathbf{k}_N ; \kappa, s_N, m_j) = \int \mathrm{d}\mathbf{r} e^{i\mathbf{r}\cdot\mathbf{q}}~\psi_\kappa^{m_j}(\mathbf{r}) \overline{\Psi}^{s_N}(\mathbf{r},\mathbf{k}_N).
\end{equation}
Since in the following we consider relations between these matrices at fixed momenta $\mathbf{q}$, $\mathbf{k}_N$ and fixed $\kappa$, we drop the explicit dependence on these variables for ease of notation.
One could now perform the angular integrals and decompose $S(s_N,m_j)$ in partial-wave amplitudes.
The connection to the approximations below are made clearer without proceeding that way.

Instead, we consider the reference system where the nucleon determines the $z$-axis, and 
$\frac{1}{2}\sigma_3\chi^{\pm1/2} = \pm\frac{1}{2}\chi^{\pm1/2}$.
In this case, $m_l = 0$ in Eq.~(\ref{eq:psi_scatter_pwe}), and the scattering state has the form
\begin{equation}
\label{eq:psi_scatter}
\Psi^{s_N}(\mathbf{r}) = 
\begin{pmatrix}
G_1(r,\theta)\chi^{s_N} + 2s_N e^{i2s_N\phi} G_2(r,\theta) \chi^{-s_N} \\
F_1(r,\theta) 2s_N\chi^{s_N} + e^{i2s_N\phi} F_2(r,\theta)\chi^{-s_N}
\end{pmatrix},
\end{equation}
where $G_i, F_i$ are complex functions determined from Eqs.~(\ref{eq:psi_radial}-\ref{eq:psi_scatter_pwe}).
For a plane wave, one has $G_2 = F_2 = 0$. 
Here, this only holds in the limit of large $r$. 
For $r$ larger than the range of the nuclear potential $R$, we recover plane-wave behavior (when the Coulomb potential is neglected), i.e.:
\begin{equation}
G_1(r,\theta) = \frac{\lvert \mathbf{k}_N\rvert}{E_N+M_N} F_1(r,\theta) \quad (r> R).
\end{equation}

From this behavior of the functions $F_{1,2}$ and $G_{1,2}$, the scattering state has the form of a free Dirac spinor with spin projection along the direction of its momentum asymptotically (at large $r$). 
Since both states correspond to different final states there is no interference, 
and since the scattering states with opposite $s_N$ are orthogonal\footnote{From Eq.~(\ref{eq:psi_scatter}) the product $\left[\Psi^{s_N}(\mathbf{r})\right]^\dagger\Psi^{-s_N}(\mathbf{r})$ is a sum of four terms, each proportional to phases $e^{\pm i\phi}$. Hence, the overlap  $\int \mathrm{d}\mathbf{r}\left[\Psi^{s_N}(\mathbf{r})\right]^\dagger\Psi^{-s_N}(\mathbf{r})$ is identically zero after integration over $\phi$.} there are no spurious contributions to the matrix elements.

The states with flipped angular momentum projection can be obtained by a rotation and parity inversion which gives
\begin{equation}
\label{eq:rot_parity}
\psi^{m}(r,\theta,\phi) \propto i \gamma^5\gamma^2 \psi^{-m}(r,\theta,-\phi),
\end{equation}
up to an overall sign determined by the parity.
We then obtain a simple relation between overlap matrices for single nucleon knockout with flipped angular momenta\footnote{In the reference system considered here, one finds $\mathbf{q}\cdot\mathbf{r} = qr\left[\sin\theta_{qN}\sin\theta\cos\phi + \cos\theta_{qN}\cos\theta\right]$. The only additional $\phi$ dependence in the integral Eq.~(\ref{eq:def_Skappa}) arises from phases in the wavefunctions. The $\phi$ integrals are thus of the form $\int_0^{2\pi}\mathrm{d}\phi e^{i\left(x\cos\phi \pm n\phi \right)} = 2\pi i^n J_n(x)$, for some natural number $n$. The integral is independent of the sign. Using Eq.~(\ref{eq:rot_parity}), the integrals that enter in $S(-s_N,-m_j)$ are identical to those in $S(s_N,m_j)$ but with opposite sign of $n$.
} 
\begin{equation}
\label{eq:S_flip_all}
S(s_N,m_j) = \pm\gamma^5\gamma^2 S(-s_N, -m_j)\gamma^2\gamma^5.
\end{equation}
When the cross section is obtained from the squared current with fixed angular momentum projections $\left[J^{\mu}(s,m)\right]^* J^\nu(s,m)$ the overall sign is irrelevant. We adopt the positive sign in the following. 

Using Eq.~(\ref{eq:S_flip_all}) and the decomposition of Eq.~(\ref{eq:Fierzlike}), it is seen that the expansion coefficients with both angular momentum projections flipped are obtained by reversing the sign of the vector components ($s$, $V^\mu$, $T^{\mu\nu}$) that do not involve $\gamma^2$ and the axial components ($p$, $A^\mu$) that do\footnote{Or vice-versa depending on the arbitrary choice of sign in Eq.~(\ref{eq:S_flip_all})}.

This shows that the purely vector current, Eq.~(\ref{eq:Gamma_vector}), or purely axial current, Eq.~(\ref{eq:Gamma_axial}), conserve parity.
For example, for the vector current built from $V^\mu$ and $T^{\mu\nu}$ 
\begin{align}
    &\left[J^{i}(s_N,m_j) \right]^* J^2(s_N,m_j) = \nonumber \\
    &-\left[J^{i}(-s_N,-m_j) \right]^* J^2(-s_N,-m_j),
\end{align}
where $i\neq 2$. Thus, elements of the hadron tensor of Eqs.~(\ref{eq:LH_incl}-\ref{eq:LH_sine}) with a single out-of-plane component are identically zero: $W^{i2} = 0$ for $i\neq 2$.
The same applies to a purely axial current.
The interaction then conserves parity, since the only odd $\phi_N$-dependence in the cross section, Eq.~(\ref{eq:LH_sine}), is the $\sin\phi_N$ dependence proportional to the antisymmetric part of the lepton tensor, and hence the helicity~\cite{Sobczyk:angles}.

Further, one sees that
\begin{equation}
    S(s_N,m_j) \neq i S(-s_N, m_j)\gamma^2\gamma^5
\end{equation}
This is only generally the case when $G_2 = F_2 = 0$.
In that case one has simple relations, Eqs.~(\ref{eq:spin_flip_relations}-\ref{eq:spin_flip_end}), between the components of Eq.~(\ref{eq:Fierzlike}) with opposite $s_N$.
The reason that these relations don't generally hold for all components of Eq.~(\ref{eq:Fierzlike}) is because of the sign change of the $\phi$-dependent phases in Eq.~(\ref{eq:psi_scatter}) under the transformation of Eq.~(\ref{eq:rot_parity}).


Lastly, we note that when the scattering state and bound state are energy eigenstates of the same (Dirac) Hamiltonian with energy eigenvalues $E = T_N$ and $E = E_\kappa$ respectively.
That is, when both states satisfy $E\psi = \left[\gamma_i\nabla^i - M_N + U(r)\right]\psi$ for the same  potential $U(r)$ we have the following relation between the components of $V^\mu(s_N,m_j)$ in Eq.~(\ref{eq:Fierzlike}):
\begin{equation}
(T_N + E_\kappa) V^0(s_N,m_j) = \mathbf{q}_i V^i(s_N,m_j). 
\label{eq:CVC_Vcomp}
\end{equation}
When this is the case, the vector current of Eq.~(\ref{eq:currentVTAP}) satisfies the Ward identity $Q \cdot J = 0$, in the absence of nuclear recoil, when $\omega = T_N + E_\kappa$.
The tensor contribution in Eq.~(\ref{eq:currentVTAP}) satisfies the Ward identity by construction.

It is clear that a model with static central potentials cannot account for recoil.
However, the nuclear recoil, $|\np_m|^2/(2M_B)$, is a small correction for medium and large nuclei.
In the relativistic mean field (RMF) calculations of Ref.~\cite{Gonzalez-Jimenez:2019ejf} initial and final-state are obtained in the same potential, such that Eq.~(\ref{eq:CVC_Vcomp}) is satisfied. In the EDRMF, the final-state potential is energy-dependent, and at large nucleon energies $200~\mathrm{MeV}\lesssim T_N$ vector current conservation is expected to be mildly violated.
The conservation of vector current can be imposed by enforcing $\mathbf{q}_iV^i \equiv \omega V^0$.

\subsection{Positive energy states and non-relativistic wavefunctions}
\label{sec:nonrel}
The matrices $S(\kappa,s_N,m_j)$ may also be determined for typical non-relativistic treatments of the wavefunctions.
We illustrate this here and relate the non-relativistic treatment to retaining only the positive-energy components of relativistic wavefunctions.

In the non-relativistic case, single-particle states may be written in terms of angular-momentum states represented by the two-component wavefunctions
\begin{equation}
\label{eq:def_psi_nr}
\psi_{\kappa,~(n.r.)}^{m_j}(\mathbf{r}) = \tilde{g}(r) \Phi_\kappa^{m_j}\left(\Omega_r\right),
\end{equation}
where the spin-spherical harmonics are given explicitly in Eq.~(\ref{eq:SSH}).
To determine a set of operators that act on such states, a non-relativistic reduction of the free-nucleon current evaluated with Dirac spinors is performed.
For example, $\overline{u}(\mathbf{p}^\prime) \Gamma^\mu u(\mathbf{p})$ may be expanded in powers of momenta and written in two-by-two form in terms of momenta and Pauli-matrices.

To make the connection to such a calculation, we examine the positive-energy projection of the four-component wavefunctions.
This is most readily done in momentum space.
The angular momentum states have the same functional form as in coordinate space
\begin{align}
\psi_\kappa^{m}(\mathbf{p}) &= \frac{1}{(2\pi)^{3/2}} \int \mathrm{d}\mathbf{r} e^{-i\mathbf{r}\cdot\mathbf{p}} \psi_\kappa^{m}(\mathbf{r}) \nonumber \\
&= 
\begin{pmatrix}
G_\kappa(p) \Phi_\kappa^{m}\left(\Omega_p \right) \\
-F_\kappa(p) \Phi_{-\kappa}^{m}\left(\Omega_p \right)
\end{pmatrix}.
\end{align}
The radial wavefunctions in momentum space, denoted by capitals, are obtained by a Hankel transform of their coordinate space counterparts~\cite{Caballero:1997gc}.
The projection on positive energy is then readily obtained, making use of $\Phi^{m}_{-\kappa}(\Omega_p) = -\bm{\sigma}\cdot\hat{\mathbf{p}}\, \Phi_{\kappa}^{m}(\Omega_p)$, to write
\begin{equation}
\label{eq:psi_positive}
\psi_{\kappa,+}^{m}(\mathbf{p}) \equiv \frac{\left(\slashed{p} + M\right)}{2M}\psi_\kappa^{m} = \tilde{G}_\kappa(\lvert \mathbf{p} \rvert) u(\mathbf{p}) \Phi_\kappa^{m}(\Omega_p),
\end{equation}
with $p^0 = E = \sqrt{\np^2 + M^2}$.
One finds a single radial wavefunction
\begin{equation}
2M\tilde{G}_\kappa(\lvert \mathbf{p} \rvert) = (E + M)G_\kappa(\lvert \mathbf{p} \rvert) - \lvert \mathbf{p} \rvert F_\kappa(\lvert \mathbf{p} \rvert).
\end{equation}
And $u(\mathbf{p}) \Phi_\kappa^m$ has the form of a positive-energy four-component Dirac spinor since:
\begin{equation}
u(\mathbf{p}) =
\begin{pmatrix}
\mathbb{1} \\
\frac{\bm\sigma\cdot\mathbf{p}}{E+M}
\end{pmatrix}.
\end{equation}
This holds for each term in the partial-wave series for the scattering state.
It is thus evident that the positive energy projection has the same structure as the non-relativistic calculation described above.
Indeed, we may write the positive-energy projected bound state in terms of non-relativistic wavefunctions of Eq.~(\ref{eq:def_psi_nr}) as
\begin{equation}
    \psi^{m}_{\kappa,+}(\mathbf{p}) = u(\mathbf{p}) \psi^{m}_{(n.r.)}(\mathbf{p}),
\end{equation}
and similarly for the scattering state $\Psi^{s}_+(\np) = u(\np) \Psi^s_{(n.r.)}(\np) $.
The current evaluated in momentum space with positive-energy states is then
\begin{align}
&J^{\mu}(s_N,m_j) = \\
&\int \mathrm{d}\mathbf{p} \left[\Psi^{s_N}_{(n.r.)}(\mathbf{p})\right]^\dagger \overline{u}(\mathbf{p}) \Gamma^\mu u(\mathbf{p} - \mathbf{q}) \psi_{\kappa,~(n.r.)}^{m_j}(\mathbf{p} - \mathbf{q}),\nonumber
\end{align}
which is identical to the non-relativistic treatment if $\overline{u}(\mathbf{p}) \Gamma^\mu u(\mathbf{p}-\mathbf{q})$ is expanded as described above.
If $\Gamma^\mu$ is taken to be momentum independent, it is clear that the momentum dependence of the non-relativistic operator is completely contained in the positive energy spinors $u(\mathbf{p})$.
As such a 4-by-4 overlap matrix $S$ may be defined with non-relativistic wavefunctions as, 
\begin{equation}
S = \int \mathrm{d}\mathbf{p} ~u(\mathbf{p} - \mathbf{q}) \psi_{\kappa,~(n.r.)}^{m_j}(\mathbf{p} - \mathbf{q}) \left[\Psi^{s_N}_{(n.r.)}(\mathbf{p})\right]^\dagger \overline{u}(\mathbf{p}).
\end{equation} 
One can perform a non-relativistic reduction of the positive energy spinors and evaluate $S$ in coordinate space, or work in momentum space to retain the full dependence.

\subsection{(Relativistic) Plane-wave impulse approximation}
\label{sec:RPWIA}
The relativistic plane-wave impulse approximation (RPWIA) gives
\begin{align}
S(\mathbf{k}_N, \mathbf{q} ; \kappa,s_N,m_j) &= \left[\int \mathrm{d}\mathbf{r} e^{-i \mathbf{p}_m\cdot\mathbf{r}} \psi_{\kappa}^{m_j}\left(\mathbf{r}\right)\right] \overline{u}(\mathbf{k}_N,s_N) \nonumber \\
&= (2\pi)^{\frac{3}{2}}\psi_\kappa^{m_j}(\mathbf{p}_m) \overline{u}(\mathbf{k}_N,s_N).
\label{eq:s_kappa_RPWIA}
\end{align}
The difference between the RPWIA and PWIA is that in the latter a positive-energy spinor, Eq.~(\ref{eq:psi_positive}), is used to describe the bound state. Implications of this difference have been discussed e.g.~in Refs.~\cite{Caballero:1997gc, Gardner:1994bh}.
Here we discuss the RPWIA\footnote{The sum over polarizations can be performed analytically to write the hadron tensor in a closed form~\cite{Gardner:1994bh, Caballero:1997gc, Nikolakopoulos:2024mjj}.
However, it is interesting to point out some properties of the overlap matrix and the differences to the more general case.}, the results also apply to the PWIA.

In the reference system considered in Eqs.~(\ref{eq:W_0_sym}-\ref{eq:W_0_asym}), the overlap matrix of Eq.~(\ref{eq:s_kappa_RPWIA}) can be represented as a completely real matrix.
With the representation of appendix~\ref{app:conventions}, this means that the coefficients $V^{\mu}, A^{\mu},$ and $T^{\mu\nu}$ of Eq.~(\ref{eq:Fierzlike}) are real when $\mu \neq 2$, and imaginary otherwise.
This implies that elements of the hadron tensor with a single out-of-plane component $W^{i2}$, $i\neq 2$ in Eq.~(\ref{eq:LH_incl}-\ref{eq:LH_sine}) are imaginary, while those without an out-of-plane component are real.
In turn, Eq.~(\ref{eq:LH_sine}) is exactly zero and there is no $\sin\phi_N$ or $\sin2\phi_N$ dependence of the cross section in the RPWIA.

Further, we have
\begin{equation}
    S(-s_N,m_j) = i S(s_N, m_j) \gamma^2\gamma^5,
\end{equation}
where we suppress the explicit dependence on $\kappa$ and the momenta, which are kept constant.
This is the case since now the functions $G_2 = F_2 = 0$ in Eq.~(\ref{eq:psi_scatter}).
This gives following simple relations between spin-flipped overlaps defined in Eq.~(\ref{eq:Fierzlike})
\begin{align}
    \label{eq:spin_flip_relations}
    s(-s_N,m) &= i A^2(s_N,m), \\
    p(-s_N,m) &= -iV^2(s_N,m), \\
    T^{ij}(-s_N,m) &= (-1)^k\epsilon^{2kji}~V^k(s_N,m), \\ 
    T^{i2}(-s_N,m) &= -iA^i(s_N,m), 
    \label{eq:spin_flip_end}
\end{align}
with $\{i,j,k\} \neq 2$. 
Here $\epsilon^{ki2j}$ is the totally antisymmetric Levi-Civita symbol with $\epsilon^{0123} = 1$.

\section{Conclusions and outlook}
\label{sec:conclusions}
We point out that under suitable assumptions the amplitudes involving the knockout of a single nucleon in the impulse approximation may be written as the trace of the product of two matrices, Eq.~(\ref{eq:Factorized}).
One contains the `nucleon dynamics', e.g. form factors, the other contains the overlap integrals of nuclear wavefunctions.
Such a factorized form is attractive for (neutrino) event generators, since the same nuclear overlaps may be used to describe electron, neutrino and BSM processes.
The factorized form makes it straightforward to consider arbitrary couplings to the nucleon, which allows for example to modify nucleon form factors in RDWIA calculations.
While in this work we write the overlap matrix in terms of four-component single particle wavefunctions, as is the case in the RDWIA, the factorized form is quite general.
We show for example that it encompasses typical calculations in non-relativistic (mean field) models.
Indeed, the latter can be mapped to the four-component form by retaining only the positive-energy projection of the single-particle wavefunctions as shown in section~\ref{sec:nonrel}.

While the factorized form of the current is quite general, in the sense that it does not rely on the adoption of a specific nuclear model, it does rely on the single-nucleon transition operator being `local'. That is the assumption that the bilinear operator $\Gamma^\mu$ of Eq.~(\ref{eq:Current_bilinear}) does not depend on $\mathbf{r}$ as discussed in section~\ref{sec:Factorized}.
This may be treated as an approximation, i.e the `asymptotic' or `local approximation', where momenta in the operator are fixed to their asymptotic values (this is of course automatically invoked when the (R)PWIA is adopted).
Corrections to this approximation may be computed for a specific model for the single-nucleon current, but it requires additional overlap matrices built from derivatives of the nuclear wavefunctions, which falls out of the scope of this work.

A general implementation of the formalism, where the traces of Eq.~(\ref{eq:Factorized}) are automated using the general form of the overlap matrix in Eq.~(\ref{eq:Fierzlike}) is highly recommended. Such efforts are currently underway in the ACHILLES event generator~\cite{Isaacson:2022cwh}.
To facilitate this, we calculate the overlaps for single-nucleon knockout from carbon and oxygen in the RDWIA using the EDRMF potential~\cite{Gonzalez-Jimenez:2019qhq}.
In section~\ref{sec:Crosssection} we give a self-contained overview of the calculation of the cross section for electron and charged-current neutrino interactions in terms of these overlap matrices.
They may be obtained along with a code that computes the cross section with arbitrary form factors from~\cite{Factorized_RDWIA_Code}.
The results from the method presented here are fully equivalent to those of the EDRMF implementation in NEUT~\cite{McKean:2025khb}, which is based on precomputed hadron tensor tables; the advantage here, is that the form factors can be changed without needing to recalculate the tables.

We produce tables for single-nucleon knockout, but these will be extended to include interactions in which additional particles are produced in the hadron system, e.g. single-pion production.
Apart from this, future work will focus on incorporating the interference between 1- and 2-body currents in single-nucleon knockout~\cite{Franco-Munoz:2022jcl,Lovato25,Casale:2025avv}, and the treatment of nucleon pairs with high relative momentum~\cite{BENHAR:RevModPhyseep}.

Lastly, we point out that one may `forget' about the nuclear model and treat the factorization in Eq.~(\ref{eq:Factorized}) as an ansatz.
The cross section can then be written in terms of a minimal set of nuclear overlaps.
Constraints on the overlaps may be derived from general principles, as done for example in section~\ref{sec:properties}, where we discuss some relations that follow from consideration of gauge-invariance and parity.
The resulting form can be used for a flexible, data-driven, (partly) model-independent parametrization of the cross section in neutrino interactions.

\begin{acknowledgments}
We thank G.~A. Miller for discussions regarding appendix C.
A.N. would like to thank K. McFarland for his talk during NuInt \#15 which inspired the true purpose of this work described in the last paragraph of the conclusion.
A.N. is supported by the Neutrino Theory Network (NTN) under Award Number DEAC02-07CH1135.
R.G.-J. was supported by projects PID2021-127098NA-I00 and RYC2022-035203-I funded by MCIN/AEI/10.13039/501100011033/FEDER and FSE+, UE;
and by ``Ayudas para Atracción de Investigadores con Alto Potencial-modalidad A'' funded by VII PPIT-US.
\end{acknowledgments}

\appendix

\section{Conventions}
\label{app:conventions}
For the calculations of overlap matrices and the argument made in section~\ref{sec:RPWIA} we use the Dirac representation
\begin{equation}
    \gamma^0 = \begin{pmatrix}
    \mathbb{1} & 0\\
    0 & - \mathbb{1}
\end{pmatrix}, \quad \gamma^{i} =  
\begin{pmatrix}
    0 & \sigma_i \\
    -\sigma_i & 0
\end{pmatrix}, \quad
\gamma^5 = \begin{pmatrix}
0 & \mathbb{1} \\
\mathbb{1} & 0
\end{pmatrix}
\end{equation}
with the Pauli matrices 
\begin{equation}
    \sigma_1 = \begin{pmatrix}
        0 & 1 \\
        1 & 0 \\
    \end{pmatrix}, \quad
        \sigma_2 = \begin{pmatrix}
        0 & -i \\
        i & 0 \\
    \end{pmatrix}, \quad
        \sigma_3 = \begin{pmatrix}
        1 & 0 \\
        0 & -1 \\
    \end{pmatrix}.
\end{equation}
Physical results in this paper are independent of this choice.

\section{Kinematics}
\label{app:kinematics}
Here we clarify in which region of phase space two solutions for the nucleon momentum are possible given the set of independent variables for the hadron system  $\omega$, $q$ and $\cos\theta_{N}$, where $\theta_{N}$ is the angle between $\nq$ and $\nk_N$.
Given these variables, Eq.~(\ref{eq:Econservation}) determines the nucleon momentum in the laboratory frame.
The solutions are 
\begin{equation}
\label{eq:kN_quad}
    \lvert \mathbf{k}_N \rvert = \frac{C q \cos\theta_{N} \pm A \sqrt{q^2\cos^2\theta_{N} M_N^2 + C^2 -A^2 M_N^2}}{q^2\cos^2\theta_{N} -A^2},
\end{equation}
where
\ba
A &\equiv& M_A +\omega\,,\nonumber\\
C &\equiv& \frac{1}{2}\bigl(q^2 + M_B^2 -A^2 -M_N^2\bigr).\nonumber
\ea
Of course, a physical solution requires $\lvert \mathbf{k}_N\rvert > 0$.
This condition excludes the solution corresponding to the positive sign in Eq.~(\ref{eq:kN_quad}) in the majority of the phase space.

The region where this second solution is positive can be better understood by considering the kinematics in the  center-of-momentum system (CMS), where $\mathbf{q}^* + \mathbf{k}_A^* = 0$.
In the CMS, there is a unique solution for the CMS momentum $\mathbf{k}^*$ such that $E_B^* = \sqrt{M_B^{2} + \mathbf{k}^{*2}}$ and $E_N^* = \sqrt{M_N^2 + \mathbf{k}^{*2}}$.
The CMS energy of the nucleon is 
\begin{equation}
\label{eq:E_N_CMS}
    E_N^* = \frac{W^2 - M_B^2 + M_N^2}{2W},
\end{equation}
where $W^2 = (\omega + M_A)^2 - q^2$.
For each scattering angle in the CMS the lab-frame momentum is determined by a boost to the lab frame.
The $z$ component of the nucleon momentum in the lab frame, with $\nq //\hat\nz$, is then given by 
\begin{equation}
    \cos\theta_{N} |\nk_N| = \gamma \left( vE_N^* + \cos\theta_{N}^* \lvert \mathbf{k}^{*}\rvert \right), \label{zboost}
\end{equation}
with $\gamma = \frac{\omega + M_A}{W}$ and $v = \frac{q}{\omega+M_A}$.
For sufficiently large velocity $v$, a solution that goes backward in the CMS ($\cos\theta_N^* < 0$) goes forward ($\cos\theta_N > 0$) in the lab system. 
One then obtains two solutions for certain forward scattering angles in the lab system. 

To determine the onset of this region such that there are two solutions for some angles, we consider the limit case $\cos\theta^*_N = -1$.
The condition is then
\begin{equation}
    v > \lvert \mathbf{k}^*\rvert/E_N^*. \label{velocity}
\end{equation}
That is: if the velocity of the CMS in the lab frame exceeds the nucleon velocity in the CMS, then two solutions are possible for certain $\cos\theta_N$.
Using Eq.~(\ref{eq:E_N_CMS}) the condition in Eq.~\eqref{velocity} can be expressed as:
\begin{equation}
    q > q_{min} = \sqrt{(\omega + M_A - M_N)^2 - M_B^2}.
\end{equation}
The equality is equivalent to
\begin{equation}
\label{eq:w_approx_Q2MB}
    \omega  = \frac{q_{min}^{2} - (\omega - E_\kappa)^2} {2M_B} + E_\kappa \approx \frac{Q^2}{2M_B} + E_\kappa.
\end{equation}
Here the nucleon kinetic energy $T_N \approx \omega - E_\kappa$ is comparable to the kinetic energy that a system with mass $M_B$ obtains in elastic scattering at the same $Q^2$. 

This region is close to the edge of the physical phase space, beyond which no solutions to Eq.~(\ref{eq:kN_quad}) exist for any scattering angle $\cos\theta_N$. 
This is where $E_N^* = M_N$, equivalent to $W = M_N + M_B$, which implies
\begin{equation}
\label{eq:q_max}
q_{max}(\omega) \equiv \sqrt{ (\omega + M_A)^2 - (M_N + M_B)^2},
\end{equation}
so that when $q > q_{max}(\omega)$ no solutions for the nucleon momentum exist for any scattering angle. 
This bound can also be derived by considering the values for $q$ so that the discriminant of eq.~\eqref{eq:kN_quad} becomes zero for $\cos\theta_N=1$. 
The bound in Eq.~(\ref{eq:q_max}) is equivalent to
\begin{equation}
    \omega = \frac{Q^2}{2M_A} + E_\kappa.
\end{equation}

\section{Possible parity-violating contributions in the symmetric hadron tensor}
\label{app:funny_tensors}
As pointed out in sec.~\ref{sec:Crosssection}, we find that the hadron tensor elements 
\begin{equation}
\label{eq:forbiddenterms}
    W^{02}_s, \quad W^{23}_s, \quad W^{12}_s
\end{equation}
which enter in the contraction of lepton and hadron tensors in Eq.~(\ref{eq:LH_sine}) are non-zero in EDRMF and RMF calculations.
This is the case only when vector and axial currents are included, since otherwise all hadron tensor elements with a single Lorentz index 2 are zero as explained in sec.~\ref{sec:properties}.

This is surprising since these terms are seemingly excluded from the general decomposition of the hadron tensor used in Refs.~\cite{Donnelly85,Moreno:2014kia}.
Indeed, in this work the symmetric hadron tensor is of the form
\begin{equation}
\label{eq:Wmunupolar}
    W^{\mu\nu}_s  = X_0 g^{\mu\nu} + \sum_i X_i \left[ V_i^\mu V_i^{\prime\nu} + V_i^\nu V_i^{\prime \mu}\right]_i,
\end{equation}
where $V_i,V_i^\prime$ are fourvectors built out of linear combinations of the set of independent fourmomenta $\{k_A, k_N, Q\}$.
The $X_i$ are functions of Lorentz scalars.
This form excludes the terms of Eq.~(\ref{eq:forbiddenterms}) since the out-of-plane components of the independent fourmomenta in the reference system considered in sec.~\ref{sec:Crosssection} are identically zero\footnote{Put otherwise: the contraction of the lepton tensor in Eq.~(\ref{eq:Leptontensor}) with Eq.~(\ref{eq:Wmunupolar}) results in terms of the form $K^\prime\cdot V_i^{(\prime)}$, $K\cdot V_i^{(\prime)}$, $K\cdot K^\prime$, and $V_i\cdot V_i^{\prime}$, none of which are proportional to components of $V_i^{(\prime)}$ orthogonal to the lepton plane.}.

To introduce (symmetric) contributions to the hadron tensor which have non-zero out of plane components one needs to use the Levi-Civita tensor.
Such symmetric tensors have been considered for example in Refs.~\cite{Hernandez:2006yg, Hernandez:2007qq}.
Following this reference, we can construct three such independent tensors.
Denoting 
\begin{equation}
    P_1^\mu = k_A^\mu, \quad P_2^\mu = k_N^\mu, \quad P_3^\mu=Q^\mu,
\end{equation}
we define
\begin{equation}
\label{eq:deffunnytensor}
    W_{Z_i}^{\mu\nu} = Z_i \left( P_i^\mu \epsilon^{\nu\alpha\beta\gamma} +P_i^\nu \epsilon^{\mu\alpha\beta\gamma}  \right)P_{1,\alpha} P_{2,\beta}P_{3,\gamma}
\end{equation}
for $i\in\{1,2,3\}$ where $Z_i$ are real scalar functions of the invariants.
These tensors are symmetric, 
but in contrast to the tensors in Eq.~(\ref{eq:Wmunupolar}) they do not behave like polar tensors under parity.
As discussed in Refs.~\cite{Hernandez:2006yg, Hernandez:2007qq} they yield parity violating contributions when contracted with the lepton tensor.
In Ref.~\cite{Sobczyk:angles}, in the context of pion production off the nucleon, it is shown that such contributions can arise when vector and axial currents with relative phases are present.
These three ingredients are of course present in the RDWIA: such terms may result in the same way from relative phases in different partial waves of Eq.~(\ref{eq:psi_scatter_pwe}) when vector and axial currents are included.

We keep the terms of Eq.~(\ref{eq:forbiddenterms}) in the general contraction of Eqs.~(\ref{eq:LH_incl}-\ref{eq:LH_sine}) since they are nonzero in our RDWIA calculations.
If these terms are forbidden by some physical principle, this would point to a fundamental deficit in the theoretical description that can provide interesting constraints on the modeling.

We provide here for completeness the kinematic dependence of the cross section arising from the tensors in Eq.~(\ref{eq:deffunnytensor}).
The contraction with the lepton tensor is
\begin{align}
    L_{\mu\nu} W_{Z_i}^{\mu\nu} = 2Z_i~(k + k^\prime)\cdot P_i \epsilon^{\alpha\beta\gamma\delta}k_\alpha k_{A,\beta} k_{N,\gamma} Q_{\delta}.
\end{align}
We evaluate the contraction in the reference system considered in sec.~\ref{sec:Crosssection}.
We define the shorthand
\begin{align}
    \mathcal{U}\sin\phi_N &\equiv 2\epsilon^{\alpha\beta\gamma\delta}k_\alpha k_{A,\beta} k_{N,\gamma} Q_{\delta} \nonumber \\
    &= -2\left(M_A \lvert \mathbf{k} \rvert \lvert \mathbf{k}^\prime \rvert \sin\theta^\prime \lvert \mathbf{k}_N\rvert\sin\theta_N \right)\sin\phi_N,
\end{align}
where $\theta^\prime$ is the angle between incoming and outgoing lepton.
One then sees that these tensors give following contributions, proportional to $\sin\phi_N$ and $\sin 2\phi_N$:
\begin{widetext}
    \begin{align}
    &L_{\mu\nu} \sum_{i=1}^{3} W_{Z_i}^{\mu\nu } 
    = \frac{1}{2}\sin 2 \phi_N~\mathcal{U} Z_2~\frac{\lvert \mathbf{k} \rvert \lvert \mathbf{k}^\prime \rvert}{q} \sin\theta^\prime \lvert \mathbf{k}_N\rvert\sin\theta_N \label{eq:sin2phiZ} \\
    &+ \sin\phi_N~\mathcal{U} \left\{ Z_1 M_A(2E_i - \omega) 
    + Z_3 \left( m^2 - m^{\prime2} \right) 
    + Z_2\left[ (2E_i-\omega)E_N 
     - \frac{\lvert\mathbf{k}_N\rvert}{\lvert \mathbf{q} \rvert}\cos\theta_N \left( \lvert\mathbf{k}\rvert^2 - \lvert \mathbf{k}^\prime \rvert^2 \right) \right]   \right\}. \label{eq:sinphiZ}
\end{align}
\end{widetext}
The tensors of Eq.~(\ref{eq:deffunnytensor}) hence only contribute to the terms in Eq.~(\ref{eq:forbiddenterms}) which are not present when the form of Eq.~(\ref{eq:Wmunupolar}) is used.
It is also clear from the explicit form of the contraction above that there is no relation between the $Z_i$ which can make both Eqs.~(\ref{eq:sin2phiZ}) and (\ref{eq:sinphiZ}) disappear\footnote{Such a relation would have to depend explicitly on lepton kinematics which is unphysical.}; this requires $Z_i = 0$.
The presence or absence of these terms may be experimentally verifiable. 
In particular the $\sin2\phi_N$ contribution, since there is no such dependence in the cross section when using Eq.~(\ref{eq:Wmunupolar})\footnote{As seen from Eq.~(\ref{eq:LH_sine}), dependence on $\sin\phi_N$ can still be generated in the absence of the funny tensors of Eq.~(\ref{eq:deffunnytensor}) in neutrino interactions through the terms involving the antisymmetric hadron tensor.}.

\bibliographystyle{apsrev4-1.bst}
\bibliography{bibliography}

\begin{thebibliography}{52}%
\makeatletter
\providecommand \@ifxundefined [1]{%
 \@ifx{#1\undefined}
}%
\providecommand \@ifnum [1]{%
 \ifnum #1\expandafter \@firstoftwo
 \else \expandafter \@secondoftwo
 \fi
}%
\providecommand \@ifx [1]{%
 \ifx #1\expandafter \@firstoftwo
 \else \expandafter \@secondoftwo
 \fi
}%
\providecommand \natexlab [1]{#1}%
\providecommand \enquote  [1]{``#1''}%
\providecommand \bibnamefont  [1]{#1}%
\providecommand \bibfnamefont [1]{#1}%
\providecommand \citenamefont [1]{#1}%
\providecommand \href@noop [0]{\@secondoftwo}%
\providecommand \href [0]{\begingroup \@sanitize@url \@href}%
\providecommand \@href[1]{\@@startlink{#1}\@@href}%
\providecommand \@@href[1]{\endgroup#1\@@endlink}%
\providecommand \@sanitize@url [0]{\catcode `\\12\catcode `\$12\catcode
  `\&12\catcode `\#12\catcode `\^12\catcode `\_12\catcode `\%12\relax}%
\providecommand \@@startlink[1]{}%
\providecommand \@@endlink[0]{}%
\providecommand \url  [0]{\begingroup\@sanitize@url \@url }%
\providecommand \@url [1]{\endgroup\@href {#1}{\urlprefix }}%
\providecommand \urlprefix  [0]{URL }%
\providecommand \Eprint [0]{\href }%
\providecommand \doibase [0]{http://dx.doi.org/}%
\providecommand \selectlanguage [0]{\@gobble}%
\providecommand \bibinfo  [0]{\@secondoftwo}%
\providecommand \bibfield  [0]{\@secondoftwo}%
\providecommand \translation [1]{[#1]}%
\providecommand \BibitemOpen [0]{}%
\providecommand \bibitemStop [0]{}%
\providecommand \bibitemNoStop [0]{.\EOS\space}%
\providecommand \EOS [0]{\spacefactor3000\relax}%
\providecommand \BibitemShut  [1]{\csname bibitem#1\endcsname}%
\let\auto@bib@innerbib\@empty
\bibitem [{\citenamefont {Alvarez-Ruso}\ \emph {et~al.}(2018)\citenamefont
  {Alvarez-Ruso}, \citenamefont {Sajjad Athar}, \citenamefont {Barbaro},
  \citenamefont {Cherdack}, \citenamefont {Christy}, \citenamefont {Coloma},
  \citenamefont {Donnelly}, \citenamefont {Dytman}, \citenamefont
  {de Gouv\^{e}a}, \citenamefont {Hill}, \citenamefont {Huber}, \citenamefont
  {Jachowicz}, \citenamefont {Katori}, \citenamefont {Kronfeld}, \citenamefont
  {Mahn}, \citenamefont {Martini}, \citenamefont {Morf\'{i}n}, \citenamefont
  {Nieves}, \citenamefont {Perdue}, \citenamefont {Petti}, \citenamefont
  {Richards}, \citenamefont {S\'{a}nchez}, \citenamefont {Sato}, \citenamefont
  {Sobczyk},\ and\ \citenamefont {Zeller}}]{NUSTECWP}%
  \BibitemOpen
  \bibfield  {author} {\bibinfo {author} {\bibfnamefont {L.}~\bibnamefont
  {Alvarez-Ruso}}, \bibinfo {author} {\bibfnamefont {M.}~\bibnamefont
  {Sajjad Athar}}, \bibinfo {author} {\bibfnamefont {M.}~\bibnamefont
  {Barbaro}}, \bibinfo {author} {\bibfnamefont {D.}~\bibnamefont {Cherdack}},
  \bibinfo {author} {\bibfnamefont {M.}~\bibnamefont {Christy}}, \bibinfo
  {author} {\bibfnamefont {P.}~\bibnamefont {Coloma}}, \bibinfo {author}
  {\bibfnamefont {T.}~\bibnamefont {Donnelly}}, \bibinfo {author}
  {\bibfnamefont {S.}~\bibnamefont {Dytman}}, \bibinfo {author} {\bibfnamefont
  {A.}~\bibnamefont {de Gouv\^{e}a}}, \bibinfo {author} {\bibfnamefont
  {R.}~\bibnamefont {Hill}}, \bibinfo {author} {\bibfnamefont {P.}~\bibnamefont
  {Huber}}, \bibinfo {author} {\bibfnamefont {N.}~\bibnamefont {Jachowicz}},
  \bibinfo {author} {\bibfnamefont {T.}~\bibnamefont {Katori}}, \bibinfo
  {author} {\bibfnamefont {A.}~\bibnamefont {Kronfeld}}, \bibinfo {author}
  {\bibfnamefont {K.}~\bibnamefont {Mahn}}, \bibinfo {author} {\bibfnamefont
  {M.}~\bibnamefont {Martini}}, \bibinfo {author} {\bibfnamefont
  {J.}~\bibnamefont {Morf\'{i}n}}, \bibinfo {author} {\bibfnamefont
  {J.}~\bibnamefont {Nieves}}, \bibinfo {author} {\bibfnamefont
  {G.}~\bibnamefont {Perdue}}, \bibinfo {author} {\bibfnamefont
  {R.}~\bibnamefont {Petti}}, \bibinfo {author} {\bibfnamefont
  {D.}~\bibnamefont {Richards}}, \bibinfo {author} {\bibfnamefont
  {F.}~\bibnamefont {S\'{a}nchez}}, \bibinfo {author} {\bibfnamefont
  {T.}~\bibnamefont {Sato}}, \bibinfo {author} {\bibfnamefont {J.}~\bibnamefont
  {Sobczyk}}, \ and\ \bibinfo {author} {\bibfnamefont {G.}~\bibnamefont
  {Zeller}},\ }\href
  {http://www.sciencedirect.com/science/article/pii/S0146641018300061}
  {\bibfield  {journal} {\bibinfo  {journal} {Progress in Particle and Nuclear
  Physics}\ }\textbf {\bibinfo {volume} {100}},\ \bibinfo {pages} {1} (\bibinfo
  {year} {2018})}\BibitemShut {NoStop}%
\bibitem [{\citenamefont {Katori}\ and\ \citenamefont
  {Martini}(2018)}]{KatoriMartinireview}%
  \BibitemOpen
  \bibfield  {author} {\bibinfo {author} {\bibfnamefont {T.}~\bibnamefont
  {Katori}}\ and\ \bibinfo {author} {\bibfnamefont {M.}~\bibnamefont
  {Martini}},\ }\href {\doibase 10.1088/1361-6471/aa8bf7} {\bibfield  {journal}
  {\bibinfo  {journal} {J. Phys.}\ }\textbf {\bibinfo {volume} {G45}},\
  \bibinfo {pages} {013001} (\bibinfo {year} {2018})},\ \Eprint
  {http://arxiv.org/abs/1611.07770} {arXiv:1611.07770 [hep-ph]} \BibitemShut
  {NoStop}%
\bibitem [{\citenamefont {Boffi}\ \emph {et~al.}(1993)\citenamefont {Boffi},
  \citenamefont {Giusti},\ and\ \citenamefont {Pacati}}]{Boffi93}%
  \BibitemOpen
  \bibfield  {author} {\bibinfo {author} {\bibfnamefont {S.}~\bibnamefont
  {Boffi}}, \bibinfo {author} {\bibfnamefont {C.}~\bibnamefont {Giusti}}, \
  and\ \bibinfo {author} {\bibfnamefont {F.~D.}\ \bibnamefont {Pacati}},\
  }\href@noop {} {\bibfield  {journal} {\bibinfo  {journal} {Phys. Rep.}\
  }\textbf {\bibinfo {volume} {226}},\ \bibinfo {pages} {1} (\bibinfo {year}
  {1993})}\BibitemShut {NoStop}%
\bibitem [{\citenamefont {Nikolakopoulos}\ \emph {et~al.}(2019)\citenamefont
  {Nikolakopoulos}, \citenamefont {Jachowicz}, \citenamefont {Van~Dessel},
  \citenamefont {Niewczas}, \citenamefont {Gonz{\'a}lez-Jim{\'e}nez},
  \citenamefont {Ud{\'\i}as},\ and\ \citenamefont
  {Pandey}}]{Nikolakopoulos:2019qcr}%
  \BibitemOpen
  \bibfield  {author} {\bibinfo {author} {\bibfnamefont {A.}~\bibnamefont
  {Nikolakopoulos}}, \bibinfo {author} {\bibfnamefont {N.}~\bibnamefont
  {Jachowicz}}, \bibinfo {author} {\bibfnamefont {N.}~\bibnamefont
  {Van~Dessel}}, \bibinfo {author} {\bibfnamefont {K.}~\bibnamefont
  {Niewczas}}, \bibinfo {author} {\bibfnamefont {R.}~\bibnamefont
  {Gonz{\'a}lez-Jim{\'e}nez}}, \bibinfo {author} {\bibfnamefont {J.~M.}\
  \bibnamefont {Ud{\'\i}as}}, \ and\ \bibinfo {author} {\bibfnamefont
  {V.}~\bibnamefont {Pandey}},\ }\href {\doibase
  10.1103/PhysRevLett.123.052501} {\bibfield  {journal} {\bibinfo  {journal}
  {Phys. Rev. Lett.}\ }\textbf {\bibinfo {volume} {123}},\ \bibinfo {pages}
  {052501} (\bibinfo {year} {2019})},\ \Eprint
  {http://arxiv.org/abs/1901.08050} {arXiv:1901.08050 [nucl-th]} \BibitemShut
  {NoStop}%
\bibitem [{\citenamefont {Gonz{\'a}lez-Jim{\'e}nez}\ \emph
  {et~al.}(2019)\citenamefont {Gonz{\'a}lez-Jim{\'e}nez}, \citenamefont
  {Nikolakopoulos}, \citenamefont {Jachowicz},\ and\ \citenamefont
  {Ud{\'\i}as}}]{Gonzalez-Jimenez:2019qhq}%
  \BibitemOpen
  \bibfield  {author} {\bibinfo {author} {\bibfnamefont {R.}~\bibnamefont
  {Gonz{\'a}lez-Jim{\'e}nez}}, \bibinfo {author} {\bibfnamefont
  {A.}~\bibnamefont {Nikolakopoulos}}, \bibinfo {author} {\bibfnamefont
  {N.}~\bibnamefont {Jachowicz}}, \ and\ \bibinfo {author} {\bibfnamefont
  {J.~M.}\ \bibnamefont {Ud{\'\i}as}},\ }\href {\doibase
  10.1103/PhysRevC.100.045501} {\bibfield  {journal} {\bibinfo  {journal}
  {Phys. Rev. C}\ }\textbf {\bibinfo {volume} {100}},\ \bibinfo {pages}
  {045501} (\bibinfo {year} {2019})},\ \Eprint
  {http://arxiv.org/abs/1904.10696} {arXiv:1904.10696 [nucl-th]} \BibitemShut
  {NoStop}%
\bibitem [{\citenamefont {Serot}(1981)}]{SEROT1981263}%
  \BibitemOpen
  \bibfield  {author} {\bibinfo {author} {\bibfnamefont {B.~D.}\ \bibnamefont
  {Serot}},\ }\href {\doibase https://doi.org/10.1016/0370-2693(81)90826-1}
  {\bibfield  {journal} {\bibinfo  {journal} {Physics Letters B}\ }\textbf
  {\bibinfo {volume} {107}},\ \bibinfo {pages} {263} (\bibinfo {year}
  {1981})}\BibitemShut {NoStop}%
\bibitem [{\citenamefont {Nikolakopoulos}\ \emph {et~al.}(2020)\citenamefont
  {Nikolakopoulos}, \citenamefont {Jachowicz}, \citenamefont
  {Gonz\'alez-Jim\'enez}, \citenamefont {Ud\'\i{}as}, \citenamefont
  {Niewczas},\ and\ \citenamefont {Pandey}}]{Nikolakopoulos:2020fti}%
  \BibitemOpen
  \bibfield  {author} {\bibinfo {author} {\bibfnamefont {A.}~\bibnamefont
  {Nikolakopoulos}}, \bibinfo {author} {\bibfnamefont {N.}~\bibnamefont
  {Jachowicz}}, \bibinfo {author} {\bibfnamefont {R.}~\bibnamefont
  {Gonz\'alez-Jim\'enez}}, \bibinfo {author} {\bibfnamefont {J.~M.}\
  \bibnamefont {Ud\'\i{}as}}, \bibinfo {author} {\bibfnamefont
  {K.}~\bibnamefont {Niewczas}}, \ and\ \bibinfo {author} {\bibfnamefont
  {V.}~\bibnamefont {Pandey}},\ }\href {\doibase 10.22323/1.369.0048}
  {\bibfield  {journal} {\bibinfo  {journal} {PoS}\ }\textbf {\bibinfo {volume}
  {NuFact2019}},\ \bibinfo {pages} {048} (\bibinfo {year} {2020})}\BibitemShut
  {NoStop}%
\bibitem [{\citenamefont {Serot}\ and\ \citenamefont
  {Walecka}(1997)}]{Serot:1997xg}%
  \BibitemOpen
  \bibfield  {author} {\bibinfo {author} {\bibfnamefont {B.~D.}\ \bibnamefont
  {Serot}}\ and\ \bibinfo {author} {\bibfnamefont {J.~D.}\ \bibnamefont
  {Walecka}},\ }\href {\doibase 10.1142/S0218301397000299} {\bibfield
  {journal} {\bibinfo  {journal} {Int. J. Mod. Phys. E}\ }\textbf {\bibinfo
  {volume} {6}},\ \bibinfo {pages} {515} (\bibinfo {year} {1997})},\ \Eprint
  {http://arxiv.org/abs/nucl-th/9701058} {arXiv:nucl-th/9701058} \BibitemShut
  {NoStop}%
\bibitem [{\citenamefont {Gonz\'alez-Jim\'enez}\ \emph
  {et~al.}(2013)\citenamefont {Gonz\'alez-Jim\'enez}, \citenamefont
  {Caballero}, \citenamefont {Meucci}, \citenamefont {Giusti}, \citenamefont
  {Barbaro}, \citenamefont {Ivanov},\ and\ \citenamefont
  {Ud\'{\i}as}}]{Gonzalez-Jimenez13c}%
  \BibitemOpen
  \bibfield  {author} {\bibinfo {author} {\bibfnamefont {R.}~\bibnamefont
  {Gonz\'alez-Jim\'enez}}, \bibinfo {author} {\bibfnamefont {J.~A.}\
  \bibnamefont {Caballero}}, \bibinfo {author} {\bibfnamefont {A.}~\bibnamefont
  {Meucci}}, \bibinfo {author} {\bibfnamefont {C.}~\bibnamefont {Giusti}},
  \bibinfo {author} {\bibfnamefont {M.~B.}\ \bibnamefont {Barbaro}}, \bibinfo
  {author} {\bibfnamefont {M.~V.}\ \bibnamefont {Ivanov}}, \ and\ \bibinfo
  {author} {\bibfnamefont {J.~M.}\ \bibnamefont {Ud\'{\i}as}},\ }\href
  {\doibase 10.1103/PhysRevC.88.025502} {\bibfield  {journal} {\bibinfo
  {journal} {Phys. Rev. C}\ }\textbf {\bibinfo {volume} {88}},\ \bibinfo
  {pages} {025502} (\bibinfo {year} {2013})}\BibitemShut {NoStop}%
\bibitem [{\citenamefont {Gonz\'alez-Jim\'enez}\ \emph
  {et~al.}(2022)\citenamefont {Gonz\'alez-Jim\'enez}, \citenamefont {Barbaro},
  \citenamefont {Caballero}, \citenamefont {Donnelly}, \citenamefont
  {Jachowicz}, \citenamefont {Megias}, \citenamefont {Niewczas}, \citenamefont
  {Nikolakopoulos}, \citenamefont {Van~Orden},\ and\ \citenamefont
  {Ud\'{\i}as}}]{Gonzalez-Jimenez:2021ohu}%
  \BibitemOpen
  \bibfield  {author} {\bibinfo {author} {\bibfnamefont {R.}~\bibnamefont
  {Gonz\'alez-Jim\'enez}}, \bibinfo {author} {\bibfnamefont {M.~B.}\
  \bibnamefont {Barbaro}}, \bibinfo {author} {\bibfnamefont {J.~A.}\
  \bibnamefont {Caballero}}, \bibinfo {author} {\bibfnamefont {T.~W.}\
  \bibnamefont {Donnelly}}, \bibinfo {author} {\bibfnamefont {N.}~\bibnamefont
  {Jachowicz}}, \bibinfo {author} {\bibfnamefont {G.~D.}\ \bibnamefont
  {Megias}}, \bibinfo {author} {\bibfnamefont {K.}~\bibnamefont {Niewczas}},
  \bibinfo {author} {\bibfnamefont {A.}~\bibnamefont {Nikolakopoulos}},
  \bibinfo {author} {\bibfnamefont {J.~W.}\ \bibnamefont {Van~Orden}}, \ and\
  \bibinfo {author} {\bibfnamefont {J.~M.}\ \bibnamefont {Ud\'{\i}as}},\ }\href
  {\doibase 10.1103/PhysRevC.105.025502} {\bibfield  {journal} {\bibinfo
  {journal} {Phys. Rev. C}\ }\textbf {\bibinfo {volume} {105}},\ \bibinfo
  {pages} {025502} (\bibinfo {year} {2022})}\BibitemShut {NoStop}%
\bibitem [{\citenamefont {Franco-Patino}\ \emph {et~al.}(2022)\citenamefont
  {Franco-Patino}, \citenamefont {Gonz\'alez-Jim\'enez}, \citenamefont {Dolan},
  \citenamefont {Barbaro}, \citenamefont {Caballero}, \citenamefont {Megias},\
  and\ \citenamefont {Udias}}]{Franco-Patino:2022tvv}%
  \BibitemOpen
  \bibfield  {author} {\bibinfo {author} {\bibfnamefont {J.~M.}\ \bibnamefont
  {Franco-Patino}}, \bibinfo {author} {\bibfnamefont {R.}~\bibnamefont
  {Gonz\'alez-Jim\'enez}}, \bibinfo {author} {\bibfnamefont {S.}~\bibnamefont
  {Dolan}}, \bibinfo {author} {\bibfnamefont {M.~B.}\ \bibnamefont {Barbaro}},
  \bibinfo {author} {\bibfnamefont {J.~A.}\ \bibnamefont {Caballero}}, \bibinfo
  {author} {\bibfnamefont {G.~D.}\ \bibnamefont {Megias}}, \ and\ \bibinfo
  {author} {\bibfnamefont {J.~M.}\ \bibnamefont {Udias}},\ }\href@noop {} {\
  (\bibinfo {year} {2022})},\ \Eprint {http://arxiv.org/abs/2207.02086}
  {arXiv:2207.02086 [nucl-th]} \BibitemShut {NoStop}%
\bibitem [{\citenamefont {Nikolakopoulos}\ \emph {et~al.}(2022)\citenamefont
  {Nikolakopoulos}, \citenamefont {Gonz\'alez-Jim\'enez}, \citenamefont
  {Jachowicz}, \citenamefont {Niewczas}, \citenamefont {S\'anchez},\ and\
  \citenamefont {Ud\'\i{}as}}]{Nikolakopoulos:2022qkq}%
  \BibitemOpen
  \bibfield  {author} {\bibinfo {author} {\bibfnamefont {A.}~\bibnamefont
  {Nikolakopoulos}}, \bibinfo {author} {\bibfnamefont {R.}~\bibnamefont
  {Gonz\'alez-Jim\'enez}}, \bibinfo {author} {\bibfnamefont {N.}~\bibnamefont
  {Jachowicz}}, \bibinfo {author} {\bibfnamefont {K.}~\bibnamefont {Niewczas}},
  \bibinfo {author} {\bibfnamefont {F.}~\bibnamefont {S\'anchez}}, \ and\
  \bibinfo {author} {\bibfnamefont {J.~M.}\ \bibnamefont {Ud\'\i{}as}},\ }\href
  {\doibase 10.1103/PhysRevC.105.054603} {\bibfield  {journal} {\bibinfo
  {journal} {Phys. Rev. C}\ }\textbf {\bibinfo {volume} {105}},\ \bibinfo
  {pages} {054603} (\bibinfo {year} {2022})},\ \Eprint
  {http://arxiv.org/abs/2202.01689} {arXiv:2202.01689 [nucl-th]} \BibitemShut
  {NoStop}%
\bibitem [{\citenamefont {Nikolakopoulos}\ \emph {et~al.}(2024)\citenamefont
  {Nikolakopoulos}, \citenamefont {Ershova}, \citenamefont
  {Gonz{\'a}lez-Jim{\'e}nez}, \citenamefont {Isaacson}, \citenamefont {Kelly},
  \citenamefont {Niewczas}, \citenamefont {Rocco},\ and\ \citenamefont
  {S{\'a}nchez}}]{Nikolakopoulos:2024mjj}%
  \BibitemOpen
  \bibfield  {author} {\bibinfo {author} {\bibfnamefont {A.}~\bibnamefont
  {Nikolakopoulos}}, \bibinfo {author} {\bibfnamefont {A.}~\bibnamefont
  {Ershova}}, \bibinfo {author} {\bibfnamefont {R.}~\bibnamefont
  {Gonz{\'a}lez-Jim{\'e}nez}}, \bibinfo {author} {\bibfnamefont
  {J.}~\bibnamefont {Isaacson}}, \bibinfo {author} {\bibfnamefont {A.~M.}\
  \bibnamefont {Kelly}}, \bibinfo {author} {\bibfnamefont {K.}~\bibnamefont
  {Niewczas}}, \bibinfo {author} {\bibfnamefont {N.}~\bibnamefont {Rocco}}, \
  and\ \bibinfo {author} {\bibfnamefont {F.}~\bibnamefont {S{\'a}nchez}},\
  }\href {\doibase 10.1103/PhysRevC.110.054611} {\bibfield  {journal} {\bibinfo
   {journal} {Phys. Rev. C}\ }\textbf {\bibinfo {volume} {110}},\ \bibinfo
  {pages} {054611} (\bibinfo {year} {2024})},\ \Eprint
  {http://arxiv.org/abs/2406.09244} {arXiv:2406.09244 [nucl-th]} \BibitemShut
  {NoStop}%
\bibitem [{\citenamefont {Butkevich}(2024)}]{PhysRevC.109.045502}%
  \BibitemOpen
  \bibfield  {author} {\bibinfo {author} {\bibfnamefont {A.~V.}\ \bibnamefont
  {Butkevich}},\ }\href {\doibase 10.1103/PhysRevC.109.045502} {\bibfield
  {journal} {\bibinfo  {journal} {Phys. Rev. C}\ }\textbf {\bibinfo {volume}
  {109}},\ \bibinfo {pages} {045502} (\bibinfo {year} {2024})}\BibitemShut
  {NoStop}%
\bibitem [{\citenamefont {McKean}\ \emph {et~al.}(2025)\citenamefont {McKean},
  \citenamefont {Gonz{\'a}lez-Jim{\'e}nez}, \citenamefont {Kabirnezhad},
  \citenamefont {Ud{\'\i}as},\ and\ \citenamefont {Uchida}}]{McKean:2025khb}%
  \BibitemOpen
  \bibfield  {author} {\bibinfo {author} {\bibfnamefont {J.}~\bibnamefont
  {McKean}}, \bibinfo {author} {\bibfnamefont {R.}~\bibnamefont
  {Gonz{\'a}lez-Jim{\'e}nez}}, \bibinfo {author} {\bibfnamefont
  {M.}~\bibnamefont {Kabirnezhad}}, \bibinfo {author} {\bibfnamefont {J.~M.}\
  \bibnamefont {Ud{\'\i}as}}, \ and\ \bibinfo {author} {\bibfnamefont
  {Y.}~\bibnamefont {Uchida}},\ }\href {\doibase 10.1103/f7x5-snmz} {\bibfield
  {journal} {\bibinfo  {journal} {Phys. Rev. D}\ }\textbf {\bibinfo {volume}
  {112}},\ \bibinfo {pages} {032009} (\bibinfo {year} {2025})},\ \Eprint
  {http://arxiv.org/abs/2502.10629} {arXiv:2502.10629 [hep-ex]} \BibitemShut
  {NoStop}%
\bibitem [{\citenamefont {Garc{\'\i}a-Marcos}\ \emph
  {et~al.}(2024)\citenamefont {Garc{\'\i}a-Marcos}, \citenamefont
  {Franco-Munoz}, \citenamefont {Gonz{\'a}lez-Jim{\'e}nez}, \citenamefont
  {Nikolakopoulos}, \citenamefont {Jachowicz},\ and\ \citenamefont
  {Ud{\'\i}as}}]{Garcia-Marcos:2023rnj}%
  \BibitemOpen
  \bibfield  {author} {\bibinfo {author} {\bibfnamefont {J.}~\bibnamefont
  {Garc{\'\i}a-Marcos}}, \bibinfo {author} {\bibfnamefont {T.}~\bibnamefont
  {Franco-Munoz}}, \bibinfo {author} {\bibfnamefont {R.}~\bibnamefont
  {Gonz{\'a}lez-Jim{\'e}nez}}, \bibinfo {author} {\bibfnamefont
  {A.}~\bibnamefont {Nikolakopoulos}}, \bibinfo {author} {\bibfnamefont
  {N.}~\bibnamefont {Jachowicz}}, \ and\ \bibinfo {author} {\bibfnamefont
  {J.~M.}\ \bibnamefont {Ud{\'\i}as}},\ }\href {\doibase
  10.1103/PhysRevC.109.024608} {\bibfield  {journal} {\bibinfo  {journal}
  {Phys. Rev. C}\ }\textbf {\bibinfo {volume} {109}},\ \bibinfo {pages}
  {024608} (\bibinfo {year} {2024})},\ \Eprint
  {http://arxiv.org/abs/2310.18056} {arXiv:2310.18056 [nucl-th]} \BibitemShut
  {NoStop}%
\bibitem [{\citenamefont {Nikolakopoulos}\ \emph {et~al.}(2023)\citenamefont
  {Nikolakopoulos}, \citenamefont {Gonz{\'a}lez-Jim{\'e}nez}, \citenamefont
  {Jachowicz},\ and\ \citenamefont {Ud{\'\i}as}}]{Nikolakopoulos:2022tut}%
  \BibitemOpen
  \bibfield  {author} {\bibinfo {author} {\bibfnamefont {A.}~\bibnamefont
  {Nikolakopoulos}}, \bibinfo {author} {\bibfnamefont {R.}~\bibnamefont
  {Gonz{\'a}lez-Jim{\'e}nez}}, \bibinfo {author} {\bibfnamefont
  {N.}~\bibnamefont {Jachowicz}}, \ and\ \bibinfo {author} {\bibfnamefont
  {J.~M.}\ \bibnamefont {Ud{\'\i}as}},\ }\href {\doibase
  10.1103/PhysRevD.107.053007} {\bibfield  {journal} {\bibinfo  {journal}
  {Phys. Rev. D}\ }\textbf {\bibinfo {volume} {107}},\ \bibinfo {pages}
  {053007} (\bibinfo {year} {2023})},\ \Eprint
  {http://arxiv.org/abs/2210.12144} {arXiv:2210.12144 [nucl-th]} \BibitemShut
  {NoStop}%
\bibitem [{\citenamefont {Gonz{\'a}lez-Jim{\'e}nez}\ \emph
  {et~al.}(2020)\citenamefont {Gonz{\'a}lez-Jim{\'e}nez}, \citenamefont
  {Barbaro}, \citenamefont {Caballero}, \citenamefont {Donnelly}, \citenamefont
  {Jachowicz}, \citenamefont {Megias}, \citenamefont {Niewczas}, \citenamefont
  {Nikolakopoulos},\ and\ \citenamefont
  {Ud{\'\i}as}}]{Gonzalez-Jimenez:2019ejf}%
  \BibitemOpen
  \bibfield  {author} {\bibinfo {author} {\bibfnamefont {R.}~\bibnamefont
  {Gonz{\'a}lez-Jim{\'e}nez}}, \bibinfo {author} {\bibfnamefont {M.~B.}\
  \bibnamefont {Barbaro}}, \bibinfo {author} {\bibfnamefont {J.~A.}\
  \bibnamefont {Caballero}}, \bibinfo {author} {\bibfnamefont {T.~W.}\
  \bibnamefont {Donnelly}}, \bibinfo {author} {\bibfnamefont {N.}~\bibnamefont
  {Jachowicz}}, \bibinfo {author} {\bibfnamefont {G.~D.}\ \bibnamefont
  {Megias}}, \bibinfo {author} {\bibfnamefont {K.}~\bibnamefont {Niewczas}},
  \bibinfo {author} {\bibfnamefont {A.}~\bibnamefont {Nikolakopoulos}}, \ and\
  \bibinfo {author} {\bibfnamefont {J.~M.}\ \bibnamefont {Ud{\'\i}as}},\ }\href
  {\doibase 10.1103/PhysRevC.101.015503} {\bibfield  {journal} {\bibinfo
  {journal} {Phys. Rev. C}\ }\textbf {\bibinfo {volume} {101}},\ \bibinfo
  {pages} {015503} (\bibinfo {year} {2020})},\ \Eprint
  {http://arxiv.org/abs/1909.07497} {arXiv:1909.07497 [nucl-th]} \BibitemShut
  {NoStop}%
\bibitem [{Fac()}]{Factorized_RDWIA_Code}%
  \BibitemOpen
  \href {https://www.github.com/alenikolak/Factorized_RDWIA/} {\enquote
  {\bibinfo {title} {github.com/alenikolak/factorized\_rdwia/},}\ }\BibitemShut
  {NoStop}%
\bibitem [{\citenamefont {Franco-Munoz}\ \emph {et~al.}(2025)\citenamefont
  {Franco-Munoz}, \citenamefont {Gonz{\'a}lez-Jim{\'e}nez},\ and\ \citenamefont
  {Ud{\'\i}as}}]{Franco-Munoz:2022jcl}%
  \BibitemOpen
  \bibfield  {author} {\bibinfo {author} {\bibfnamefont {T.}~\bibnamefont
  {Franco-Munoz}}, \bibinfo {author} {\bibfnamefont {R.}~\bibnamefont
  {Gonz{\'a}lez-Jim{\'e}nez}}, \ and\ \bibinfo {author} {\bibfnamefont {J.~M.}\
  \bibnamefont {Ud{\'\i}as}},\ }\href {\doibase 10.1088/1361-6471/ad9eca}
  {\bibfield  {journal} {\bibinfo  {journal} {J. Phys. G}\ }\textbf {\bibinfo
  {volume} {52}},\ \bibinfo {pages} {025103} (\bibinfo {year} {2025})},\
  \Eprint {http://arxiv.org/abs/2203.09996} {arXiv:2203.09996 [nucl-th]}
  \BibitemShut {NoStop}%
\bibitem [{\citenamefont {Franco-Munoz}\ \emph {et~al.}(2023)\citenamefont
  {Franco-Munoz}, \citenamefont {Garc{\'\i}a-Marcos}, \citenamefont
  {Gonz{\'a}lez-Jim{\'e}nez},\ and\ \citenamefont
  {Ud{\'\i}as}}]{Franco-Munoz:2023zoa}%
  \BibitemOpen
  \bibfield  {author} {\bibinfo {author} {\bibfnamefont {T.}~\bibnamefont
  {Franco-Munoz}}, \bibinfo {author} {\bibfnamefont {J.}~\bibnamefont
  {Garc{\'\i}a-Marcos}}, \bibinfo {author} {\bibfnamefont {R.}~\bibnamefont
  {Gonz{\'a}lez-Jim{\'e}nez}}, \ and\ \bibinfo {author} {\bibfnamefont {J.~M.}\
  \bibnamefont {Ud{\'\i}as}},\ }\href {\doibase 10.1103/PhysRevC.108.064608}
  {\bibfield  {journal} {\bibinfo  {journal} {Phys. Rev. C}\ }\textbf {\bibinfo
  {volume} {108}},\ \bibinfo {pages} {064608} (\bibinfo {year} {2023})},\
  \Eprint {http://arxiv.org/abs/2306.10823} {arXiv:2306.10823 [nucl-th]}
  \BibitemShut {NoStop}%
\bibitem [{\citenamefont {Van~Orden}\ and\ \citenamefont
  {Donnelly}(2019)}]{VanOrden:2019krz}%
  \BibitemOpen
  \bibfield  {author} {\bibinfo {author} {\bibfnamefont {J.~W.}\ \bibnamefont
  {Van~Orden}}\ and\ \bibinfo {author} {\bibfnamefont {T.~W.}\ \bibnamefont
  {Donnelly}},\ }\href {\doibase 10.1103/PhysRevC.100.044620} {\bibfield
  {journal} {\bibinfo  {journal} {Phys. Rev. C}\ }\textbf {\bibinfo {volume}
  {100}},\ \bibinfo {pages} {044620} (\bibinfo {year} {2019})},\ \Eprint
  {http://arxiv.org/abs/1908.00932} {arXiv:1908.00932 [nucl-th]} \BibitemShut
  {NoStop}%
\bibitem [{\citenamefont {Franco-Patino}\ \emph {et~al.}(2020)\citenamefont
  {Franco-Patino}, \citenamefont {Gonzalez-Rosa}, \citenamefont {Caballero},\
  and\ \citenamefont {Barbaro}}]{PhysRevC.102.064626}%
  \BibitemOpen
  \bibfield  {author} {\bibinfo {author} {\bibfnamefont {J.~M.}\ \bibnamefont
  {Franco-Patino}}, \bibinfo {author} {\bibfnamefont {J.}~\bibnamefont
  {Gonzalez-Rosa}}, \bibinfo {author} {\bibfnamefont {J.~A.}\ \bibnamefont
  {Caballero}}, \ and\ \bibinfo {author} {\bibfnamefont {M.~B.}\ \bibnamefont
  {Barbaro}},\ }\href {\doibase 10.1103/PhysRevC.102.064626} {\bibfield
  {journal} {\bibinfo  {journal} {Phys. Rev. C}\ }\textbf {\bibinfo {volume}
  {102}},\ \bibinfo {pages} {064626} (\bibinfo {year} {2020})}\BibitemShut
  {NoStop}%
\bibitem [{\citenamefont {Dickhoff}\ and\ \citenamefont
  {Van~Neck}(2008)}]{dickhoff2008many}%
  \BibitemOpen
  \bibfield  {author} {\bibinfo {author} {\bibfnamefont {W.~H.}\ \bibnamefont
  {Dickhoff}}\ and\ \bibinfo {author} {\bibfnamefont {D.~V.}\ \bibnamefont
  {Van~Neck}},\ }\href@noop {} {\emph {\bibinfo {title} {Many-body theory
  exposed! Propagator description of quantum mechanics in many-body systems}}}\
  (\bibinfo  {publisher} {World Scientific Publishing Company},\ \bibinfo
  {year} {2008})\BibitemShut {NoStop}%
\bibitem [{\citenamefont {Rocco}\ \emph {et~al.}(2019)\citenamefont {Rocco},
  \citenamefont {Nakamura}, \citenamefont {Lee},\ and\ \citenamefont
  {Lovato}}]{Rocco:2019gfb}%
  \BibitemOpen
  \bibfield  {author} {\bibinfo {author} {\bibfnamefont {N.}~\bibnamefont
  {Rocco}}, \bibinfo {author} {\bibfnamefont {S.~X.}\ \bibnamefont {Nakamura}},
  \bibinfo {author} {\bibfnamefont {T.~S.~H.}\ \bibnamefont {Lee}}, \ and\
  \bibinfo {author} {\bibfnamefont {A.}~\bibnamefont {Lovato}},\ }\href
  {\doibase 10.1103/PhysRevC.100.045503} {\bibfield  {journal} {\bibinfo
  {journal} {Phys. Rev. C}\ }\textbf {\bibinfo {volume} {100}},\ \bibinfo
  {pages} {045503} (\bibinfo {year} {2019})},\ \Eprint
  {http://arxiv.org/abs/1907.01093} {arXiv:1907.01093 [nucl-th]} \BibitemShut
  {NoStop}%
\bibitem [{\citenamefont {Kopp}\ \emph {et~al.}(2024)\citenamefont {Kopp},
  \citenamefont {Rocco},\ and\ \citenamefont {Tabrizi}}]{Kopp:2024yvh}%
  \BibitemOpen
  \bibfield  {author} {\bibinfo {author} {\bibfnamefont {J.}~\bibnamefont
  {Kopp}}, \bibinfo {author} {\bibfnamefont {N.}~\bibnamefont {Rocco}}, \ and\
  \bibinfo {author} {\bibfnamefont {Z.}~\bibnamefont {Tabrizi}},\ }\href
  {\doibase 10.1007/JHEP08(2024)187} {\bibfield  {journal} {\bibinfo  {journal}
  {JHEP}\ }\textbf {\bibinfo {volume} {08}},\ \bibinfo {pages} {187} (\bibinfo
  {year} {2024})},\ \Eprint {http://arxiv.org/abs/2401.07902} {arXiv:2401.07902
  [hep-ph]} \BibitemShut {NoStop}%
\bibitem [{\citenamefont {Caballero}\ \emph {et~al.}(1998)\citenamefont
  {Caballero}, \citenamefont {Donnelly}, \citenamefont {Moya~de Guerra},\ and\
  \citenamefont {Udias}}]{Caballero:1997gc}%
  \BibitemOpen
  \bibfield  {author} {\bibinfo {author} {\bibfnamefont {J.~A.}\ \bibnamefont
  {Caballero}}, \bibinfo {author} {\bibfnamefont {T.~W.}\ \bibnamefont
  {Donnelly}}, \bibinfo {author} {\bibfnamefont {E.}~\bibnamefont {Moya~de
  Guerra}}, \ and\ \bibinfo {author} {\bibfnamefont {J.~M.}\ \bibnamefont
  {Udias}},\ }\href {\doibase 10.1016/S0375-9474(97)00817-8} {\bibfield
  {journal} {\bibinfo  {journal} {Nucl. Phys. A}\ }\textbf {\bibinfo {volume}
  {632}},\ \bibinfo {pages} {323} (\bibinfo {year} {1998})},\ \Eprint
  {http://arxiv.org/abs/nucl-th/9710038} {arXiv:nucl-th/9710038} \BibitemShut
  {NoStop}%
\bibitem [{\citenamefont {Ud\'{\i}as}\ \emph {et~al.}(1993)\citenamefont
  {Ud\'{\i}as}, \citenamefont {Sarriguren}, \citenamefont {Moya~de Guerra},
  \citenamefont {Garrido},\ and\ \citenamefont {Caballero}}]{Udias93}%
  \BibitemOpen
  \bibfield  {author} {\bibinfo {author} {\bibfnamefont {J.~M.}\ \bibnamefont
  {Ud\'{\i}as}}, \bibinfo {author} {\bibfnamefont {P.}~\bibnamefont
  {Sarriguren}}, \bibinfo {author} {\bibfnamefont {E.}~\bibnamefont {Moya~de
  Guerra}}, \bibinfo {author} {\bibfnamefont {E.}~\bibnamefont {Garrido}}, \
  and\ \bibinfo {author} {\bibfnamefont {J.~A.}\ \bibnamefont {Caballero}},\
  }\href {\doibase 10.1103/PhysRevC.48.2731} {\bibfield  {journal} {\bibinfo
  {journal} {Phys. Rev. C}\ }\textbf {\bibinfo {volume} {48}},\ \bibinfo
  {pages} {2731} (\bibinfo {year} {1993})}\BibitemShut {NoStop}%
\bibitem [{\citenamefont {Jeschonnek}\ and\ \citenamefont
  {Donnelly}(1998)}]{Jeschonnek:1997dm}%
  \BibitemOpen
  \bibfield  {author} {\bibinfo {author} {\bibfnamefont {S.}~\bibnamefont
  {Jeschonnek}}\ and\ \bibinfo {author} {\bibfnamefont {T.~W.}\ \bibnamefont
  {Donnelly}},\ }\href {\doibase 10.1103/PhysRevC.57.2438} {\bibfield
  {journal} {\bibinfo  {journal} {Phys. Rev. C}\ }\textbf {\bibinfo {volume}
  {57}},\ \bibinfo {pages} {2438} (\bibinfo {year} {1998})},\ \Eprint
  {http://arxiv.org/abs/nucl-th/9711014} {arXiv:nucl-th/9711014} \BibitemShut
  {NoStop}%
\bibitem [{\citenamefont {Walecka}(2012)}]{walecka2012semileptonic}%
  \BibitemOpen
  \bibfield  {author} {\bibinfo {author} {\bibfnamefont {J.}~\bibnamefont
  {Walecka}},\ }\href@noop {} {\bibfield  {journal} {\bibinfo  {journal} {Muon
  physics}\ }\textbf {\bibinfo {volume} {2}},\ \bibinfo {pages} {113} (\bibinfo
  {year} {2012})}\BibitemShut {NoStop}%
\bibitem [{\citenamefont {OConnell}\ \emph {et~al.}(1972)\citenamefont
  {OConnell}, \citenamefont {Donnelly},\ and\ \citenamefont
  {Walecka}}]{Waleckapaper}%
  \BibitemOpen
  \bibfield  {author} {\bibinfo {author} {\bibfnamefont {J.~S.}\ \bibnamefont
  {OConnell}}, \bibinfo {author} {\bibfnamefont {T.~W.}\ \bibnamefont
  {Donnelly}}, \ and\ \bibinfo {author} {\bibfnamefont {J.~D.}\ \bibnamefont
  {Walecka}},\ }\href {\doibase 10.1103/PhysRevC.6.719} {\bibfield  {journal}
  {\bibinfo  {journal} {Phys. Rev. C}\ }\textbf {\bibinfo {volume} {6}},\
  \bibinfo {pages} {719} (\bibinfo {year} {1972})}\BibitemShut {NoStop}%
\bibitem [{\citenamefont {Umino}\ and\ \citenamefont
  {Ud\'{i}as}(1995)}]{Udias:95}%
  \BibitemOpen
  \bibfield  {author} {\bibinfo {author} {\bibfnamefont {Y.}~\bibnamefont
  {Umino}}\ and\ \bibinfo {author} {\bibfnamefont {J.~M.}\ \bibnamefont
  {Ud\'{i}as}},\ }\href {\doibase 10.1103/PhysRevC.52.3399} {\bibfield
  {journal} {\bibinfo  {journal} {Phys. Rev. C}\ }\textbf {\bibinfo {volume}
  {52}},\ \bibinfo {pages} {3399} (\bibinfo {year} {1995})}\BibitemShut
  {NoStop}%
\bibitem [{\citenamefont {Amaro}\ \emph {et~al.}(2007)\citenamefont {Amaro},
  \citenamefont {Barbaro}, \citenamefont {Caballero}, \citenamefont
  {Donnelly},\ and\ \citenamefont {Udias}}]{Amaro07}%
  \BibitemOpen
  \bibfield  {author} {\bibinfo {author} {\bibfnamefont {J.~E.}\ \bibnamefont
  {Amaro}}, \bibinfo {author} {\bibfnamefont {M.~B.}\ \bibnamefont {Barbaro}},
  \bibinfo {author} {\bibfnamefont {J.~A.}\ \bibnamefont {Caballero}}, \bibinfo
  {author} {\bibfnamefont {T.~W.}\ \bibnamefont {Donnelly}}, \ and\ \bibinfo
  {author} {\bibfnamefont {J.~M.}\ \bibnamefont {Udias}},\ }\href {\doibase
  10.1103/PhysRevC.75.034613} {\bibfield  {journal} {\bibinfo  {journal} {Phys.
  Rev. C}\ }\textbf {\bibinfo {volume} {75}},\ \bibinfo {pages} {034613}
  (\bibinfo {year} {2007})}\BibitemShut {NoStop}%
\bibitem [{\citenamefont {Laget}(1972)}]{LAGET197281}%
  \BibitemOpen
  \bibfield  {author} {\bibinfo {author} {\bibfnamefont {J.}~\bibnamefont
  {Laget}},\ }\href {\doibase https://doi.org/10.1016/0375-9474(72)91053-6}
  {\bibfield  {journal} {\bibinfo  {journal} {Nuclear Physics A}\ }\textbf
  {\bibinfo {volume} {194}},\ \bibinfo {pages} {81} (\bibinfo {year}
  {1972})}\BibitemShut {NoStop}%
\bibitem [{\citenamefont {Vanderhaeghen}()}]{VanderhaeghenPhD}%
  \BibitemOpen
  \bibfield  {author} {\bibinfo {author} {\bibfnamefont {M.}~\bibnamefont
  {Vanderhaeghen}},\ }\emph {\bibinfo {title} {Pion production off nucleons and
  nuclei in electromagnetic nuclear reactions}},\ \href@noop {} {Ph.D.
  thesis},\ \bibinfo  {school} {Ghent University}\BibitemShut {NoStop}%
\bibitem [{\citenamefont {Li}\ \emph {et~al.}(1993)\citenamefont {Li},
  \citenamefont {Wright},\ and\ \citenamefont {Bennhold}}]{PhysRevC.48.816}%
  \BibitemOpen
  \bibfield  {author} {\bibinfo {author} {\bibfnamefont {X.}~\bibnamefont
  {Li}}, \bibinfo {author} {\bibfnamefont {L.~E.}\ \bibnamefont {Wright}}, \
  and\ \bibinfo {author} {\bibfnamefont {C.}~\bibnamefont {Bennhold}},\ }\href
  {\doibase 10.1103/PhysRevC.48.816} {\bibfield  {journal} {\bibinfo  {journal}
  {Phys. Rev. C}\ }\textbf {\bibinfo {volume} {48}},\ \bibinfo {pages} {816}
  (\bibinfo {year} {1993})}\BibitemShut {NoStop}%
\bibitem [{\citenamefont {Toker}\ and\ \citenamefont
  {Tabakin}(1983)}]{PhysRevC.28.1725}%
  \BibitemOpen
  \bibfield  {author} {\bibinfo {author} {\bibfnamefont {G.}~\bibnamefont
  {Toker}}\ and\ \bibinfo {author} {\bibfnamefont {F.}~\bibnamefont
  {Tabakin}},\ }\href {\doibase 10.1103/PhysRevC.28.1725} {\bibfield  {journal}
  {\bibinfo  {journal} {Phys. Rev. C}\ }\textbf {\bibinfo {volume} {28}},\
  \bibinfo {pages} {1725} (\bibinfo {year} {1983})}\BibitemShut {NoStop}%
\bibitem [{\citenamefont {Gonz\'alez-Jim\'enez}\ \emph
  {et~al.}(2019)\citenamefont {Gonz\'alez-Jim\'enez}, \citenamefont
  {Nikolakopoulos}, \citenamefont {Jachowicz},\ and\ \citenamefont
  {Ud\'{\i}as}}]{Gonzalez-Jimenez19}%
  \BibitemOpen
  \bibfield  {author} {\bibinfo {author} {\bibfnamefont {R.}~\bibnamefont
  {Gonz\'alez-Jim\'enez}}, \bibinfo {author} {\bibfnamefont {A.}~\bibnamefont
  {Nikolakopoulos}}, \bibinfo {author} {\bibfnamefont {N.}~\bibnamefont
  {Jachowicz}}, \ and\ \bibinfo {author} {\bibfnamefont {J.~M.}\ \bibnamefont
  {Ud\'{\i}as}},\ }\href {\doibase 10.1103/PhysRevC.100.045501} {\bibfield
  {journal} {\bibinfo  {journal} {Phys. Rev. C}\ }\textbf {\bibinfo {volume}
  {100}},\ \bibinfo {pages} {045501} (\bibinfo {year} {2019})}\BibitemShut
  {NoStop}%
\bibitem [{\citenamefont {Ud{\'i}as}(1993)}]{UdiasPhD}%
  \BibitemOpen
  \bibfield  {author} {\bibinfo {author} {\bibfnamefont {J.~M.}\ \bibnamefont
  {Ud{\'i}as}},\ }\emph {\bibinfo {title} {An{\'a}lisis Relativista del proceso
  $(e,e'p)$ en N\'ucleos Complejos}},\ \href@noop {} {Ph.D. thesis},\ \bibinfo
  {school} {Instituto de Estructura de la Materia, C.S.I.C., Madrid} (\bibinfo
  {year} {1993})\BibitemShut {NoStop}%
\bibitem [{\citenamefont {Sobczyk}\ \emph {et~al.}(2018)\citenamefont
  {Sobczyk}, \citenamefont {Hern\'andez}, \citenamefont {Nakamura},
  \citenamefont {Nieves},\ and\ \citenamefont {Sato}}]{Sobczyk:angles}%
  \BibitemOpen
  \bibfield  {author} {\bibinfo {author} {\bibfnamefont {J.~E.}\ \bibnamefont
  {Sobczyk}}, \bibinfo {author} {\bibfnamefont {E.}~\bibnamefont
  {Hern\'andez}}, \bibinfo {author} {\bibfnamefont {S.~X.}\ \bibnamefont
  {Nakamura}}, \bibinfo {author} {\bibfnamefont {J.}~\bibnamefont {Nieves}}, \
  and\ \bibinfo {author} {\bibfnamefont {T.}~\bibnamefont {Sato}},\ }\href
  {\doibase 10.1103/PhysRevD.98.073001} {\bibfield  {journal} {\bibinfo
  {journal} {Phys. Rev. D}\ }\textbf {\bibinfo {volume} {98}},\ \bibinfo
  {pages} {073001} (\bibinfo {year} {2018})}\BibitemShut {NoStop}%
\bibitem [{\citenamefont {Ivanov}\ \emph {et~al.}(2015)\citenamefont {Ivanov},
  \citenamefont {Antonov}, \citenamefont {Barbaro}, \citenamefont {Giusti},
  \citenamefont {Meucci}, \citenamefont {Caballero}, \citenamefont
  {Gonz\'alez-Jim\'enez}, \citenamefont {de~Guerra},\ and\ \citenamefont
  {Ud\'{\i}as}}]{Ivanov15}%
  \BibitemOpen
  \bibfield  {author} {\bibinfo {author} {\bibfnamefont {M.~V.}\ \bibnamefont
  {Ivanov}}, \bibinfo {author} {\bibfnamefont {A.~N.}\ \bibnamefont {Antonov}},
  \bibinfo {author} {\bibfnamefont {M.~B.}\ \bibnamefont {Barbaro}}, \bibinfo
  {author} {\bibfnamefont {C.}~\bibnamefont {Giusti}}, \bibinfo {author}
  {\bibfnamefont {A.}~\bibnamefont {Meucci}}, \bibinfo {author} {\bibfnamefont
  {J.~A.}\ \bibnamefont {Caballero}}, \bibinfo {author} {\bibfnamefont
  {R.}~\bibnamefont {Gonz\'alez-Jim\'enez}}, \bibinfo {author} {\bibfnamefont
  {E.~M.}\ \bibnamefont {de~Guerra}}, \ and\ \bibinfo {author} {\bibfnamefont
  {J.~M.}\ \bibnamefont {Ud\'{\i}as}},\ }\href {\doibase
  10.1103/PhysRevC.91.034607} {\bibfield  {journal} {\bibinfo  {journal} {Phys.
  Rev. C}\ }\textbf {\bibinfo {volume} {91}},\ \bibinfo {pages} {034607}
  (\bibinfo {year} {2015})}\BibitemShut {NoStop}%
\bibitem [{\citenamefont {González-Jiménez}\ \emph
  {et~al.}(2013)\citenamefont {González-Jiménez}, \citenamefont {Caballero},\
  and\ \citenamefont {Donnelly}}]{Gonzalez-Jimenez13}%
  \BibitemOpen
  \bibfield  {author} {\bibinfo {author} {\bibfnamefont {R.}~\bibnamefont
  {González-Jiménez}}, \bibinfo {author} {\bibfnamefont {J.}~\bibnamefont
  {Caballero}}, \ and\ \bibinfo {author} {\bibfnamefont {T.}~\bibnamefont
  {Donnelly}},\ }\href {\doibase https://doi.org/10.1016/j.physrep.2012.10.003}
  {\bibfield  {journal} {\bibinfo  {journal} {Physics Reports}\ }\textbf
  {\bibinfo {volume} {524}},\ \bibinfo {pages} {1} (\bibinfo {year}
  {2013})}\BibitemShut {NoStop}%
\bibitem [{\citenamefont {Greiner}(2000)}]{Greiner}%
  \BibitemOpen
  \bibfield  {author} {\bibinfo {author} {\bibfnamefont {W.}~\bibnamefont
  {Greiner}},\ }\href@noop {} {\emph {\bibinfo {title} {Relativistic Quantum
  Mechanics: Wave Equations}}}\ (\bibinfo  {publisher} {Springer-Verlag},\
  \bibinfo {year} {2000})\BibitemShut {NoStop}%
\bibitem [{\citenamefont {Gardner}\ and\ \citenamefont
  {Piekarewicz}(1994)}]{Gardner:1994bh}%
  \BibitemOpen
  \bibfield  {author} {\bibinfo {author} {\bibfnamefont {S.}~\bibnamefont
  {Gardner}}\ and\ \bibinfo {author} {\bibfnamefont {J.}~\bibnamefont
  {Piekarewicz}},\ }\href {\doibase 10.1103/PhysRevC.50.2822} {\bibfield
  {journal} {\bibinfo  {journal} {Phys. Rev. C}\ }\textbf {\bibinfo {volume}
  {50}},\ \bibinfo {pages} {2822} (\bibinfo {year} {1994})},\ \Eprint
  {http://arxiv.org/abs/nucl-th/9401001} {arXiv:nucl-th/9401001} \BibitemShut
  {NoStop}%
\bibitem [{\citenamefont {Isaacson}\ \emph {et~al.}(2022)\citenamefont
  {Isaacson}, \citenamefont {Jay}, \citenamefont {Lovato}, \citenamefont
  {Machado},\ and\ \citenamefont {Rocco}}]{Isaacson:2022cwh}%
  \BibitemOpen
  \bibfield  {author} {\bibinfo {author} {\bibfnamefont {J.}~\bibnamefont
  {Isaacson}}, \bibinfo {author} {\bibfnamefont {W.~I.}\ \bibnamefont {Jay}},
  \bibinfo {author} {\bibfnamefont {A.}~\bibnamefont {Lovato}}, \bibinfo
  {author} {\bibfnamefont {P.~A.~N.}\ \bibnamefont {Machado}}, \ and\ \bibinfo
  {author} {\bibfnamefont {N.}~\bibnamefont {Rocco}},\ }\href@noop {} {\
  (\bibinfo {year} {2022})},\ \Eprint {http://arxiv.org/abs/2205.06378}
  {arXiv:2205.06378 [hep-ph]} \BibitemShut {NoStop}%
\bibitem [{\citenamefont {Lovato}\ \emph {et~al.}(2025)\citenamefont {Lovato},
  \citenamefont {Rocco},\ and\ \citenamefont {Steinberg}}]{Lovato25}%
  \BibitemOpen
  \bibfield  {author} {\bibinfo {author} {\bibfnamefont {A.}~\bibnamefont
  {Lovato}}, \bibinfo {author} {\bibfnamefont {N.}~\bibnamefont {Rocco}}, \
  and\ \bibinfo {author} {\bibfnamefont {N.}~\bibnamefont {Steinberg}},\ }\href
  {\doibase 10.1103/m645-5whh} {\bibfield  {journal} {\bibinfo  {journal}
  {Phys. Rev. C}\ }\textbf {\bibinfo {volume} {112}},\ \bibinfo {pages}
  {045501} (\bibinfo {year} {2025})}\BibitemShut {NoStop}%
\bibitem [{\citenamefont {Casale}\ \emph {et~al.}(2025)\citenamefont {Casale},
  \citenamefont {Amaro}, \citenamefont {Belocchi}, \citenamefont {Barbaro},\
  and\ \citenamefont {Martini}}]{Casale:2025avv}%
  \BibitemOpen
  \bibfield  {author} {\bibinfo {author} {\bibfnamefont {P.~R.}\ \bibnamefont
  {Casale}}, \bibinfo {author} {\bibfnamefont {J.~E.}\ \bibnamefont {Amaro}},
  \bibinfo {author} {\bibfnamefont {V.}~\bibnamefont {Belocchi}}, \bibinfo
  {author} {\bibfnamefont {M.~B.}\ \bibnamefont {Barbaro}}, \ and\ \bibinfo
  {author} {\bibfnamefont {M.}~\bibnamefont {Martini}},\ }\href {\doibase
  10.1103/cp3s-sch1} {\bibfield  {journal} {\bibinfo  {journal} {Phys. Rev. C}\
  }\textbf {\bibinfo {volume} {112}},\ \bibinfo {pages} {065502} (\bibinfo
  {year} {2025})},\ \Eprint {http://arxiv.org/abs/2507.20246} {arXiv:2507.20246
  [hep-ph]} \BibitemShut {NoStop}%
\bibitem [{\citenamefont {Benhar}\ \emph {et~al.}(1993)\citenamefont {Benhar},
  \citenamefont {Pandharipande},\ and\ \citenamefont
  {Pieper}}]{BENHAR:RevModPhyseep}%
  \BibitemOpen
  \bibfield  {author} {\bibinfo {author} {\bibfnamefont {O.}~\bibnamefont
  {Benhar}}, \bibinfo {author} {\bibfnamefont {V.~R.}\ \bibnamefont
  {Pandharipande}}, \ and\ \bibinfo {author} {\bibfnamefont {S.~C.}\
  \bibnamefont {Pieper}},\ }\href {\doibase 10.1103/RevModPhys.65.817}
  {\bibfield  {journal} {\bibinfo  {journal} {Rev. Mod. Phys.}\ }\textbf
  {\bibinfo {volume} {65}},\ \bibinfo {pages} {817} (\bibinfo {year}
  {1993})}\BibitemShut {NoStop}%
\bibitem [{\citenamefont {Donnelly}(1985)}]{Donnelly85}%
  \BibitemOpen
  \bibfield  {author} {\bibinfo {author} {\bibfnamefont {T.}~\bibnamefont
  {Donnelly}},\ }\href {\doibase https://doi.org/10.1016/0146-6410(85)90008-0}
  {\bibfield  {journal} {\bibinfo  {journal} {Progress in Particle and Nuclear
  Physics}\ }\textbf {\bibinfo {volume} {13}},\ \bibinfo {pages} {183}
  (\bibinfo {year} {1985})}\BibitemShut {NoStop}%
\bibitem [{\citenamefont {Moreno}\ \emph {et~al.}(2014)\citenamefont {Moreno},
  \citenamefont {Donnelly}, \citenamefont {Van~Orden},\ and\ \citenamefont
  {Ford}}]{Moreno:2014kia}%
  \BibitemOpen
  \bibfield  {author} {\bibinfo {author} {\bibfnamefont {O.}~\bibnamefont
  {Moreno}}, \bibinfo {author} {\bibfnamefont {T.~W.}\ \bibnamefont
  {Donnelly}}, \bibinfo {author} {\bibfnamefont {J.~W.}\ \bibnamefont
  {Van~Orden}}, \ and\ \bibinfo {author} {\bibfnamefont {W.~P.}\ \bibnamefont
  {Ford}},\ }\href {\doibase 10.1103/PhysRevD.90.013014} {\bibfield  {journal}
  {\bibinfo  {journal} {Phys. Rev. D}\ }\textbf {\bibinfo {volume} {90}},\
  \bibinfo {pages} {013014} (\bibinfo {year} {2014})},\ \Eprint
  {http://arxiv.org/abs/1406.4494} {arXiv:1406.4494 [hep-th]} \BibitemShut
  {NoStop}%
\bibitem [{\citenamefont {Hernandez}\ \emph
  {et~al.}(2007{\natexlab{a}})\citenamefont {Hernandez}, \citenamefont
  {Nieves},\ and\ \citenamefont {Valverde}}]{Hernandez:2006yg}%
  \BibitemOpen
  \bibfield  {author} {\bibinfo {author} {\bibfnamefont {E.}~\bibnamefont
  {Hernandez}}, \bibinfo {author} {\bibfnamefont {J.}~\bibnamefont {Nieves}}, \
  and\ \bibinfo {author} {\bibfnamefont {M.}~\bibnamefont {Valverde}},\ }\href
  {\doibase 10.1016/j.physletb.2007.02.051} {\bibfield  {journal} {\bibinfo
  {journal} {Phys. Lett. B}\ }\textbf {\bibinfo {volume} {647}},\ \bibinfo
  {pages} {452} (\bibinfo {year} {2007}{\natexlab{a}})},\ \Eprint
  {http://arxiv.org/abs/hep-ph/0608119} {arXiv:hep-ph/0608119} \BibitemShut
  {NoStop}%
\bibitem [{\citenamefont {Hernandez}\ \emph
  {et~al.}(2007{\natexlab{b}})\citenamefont {Hernandez}, \citenamefont
  {Nieves},\ and\ \citenamefont {Valverde}}]{Hernandez:2007qq}%
  \BibitemOpen
  \bibfield  {author} {\bibinfo {author} {\bibfnamefont {E.}~\bibnamefont
  {Hernandez}}, \bibinfo {author} {\bibfnamefont {J.}~\bibnamefont {Nieves}}, \
  and\ \bibinfo {author} {\bibfnamefont {M.}~\bibnamefont {Valverde}},\ }\href
  {\doibase 10.1103/PhysRevD.76.033005} {\bibfield  {journal} {\bibinfo
  {journal} {Phys. Rev. D}\ }\textbf {\bibinfo {volume} {76}},\ \bibinfo
  {pages} {033005} (\bibinfo {year} {2007}{\natexlab{b}})},\ \Eprint
  {http://arxiv.org/abs/hep-ph/0701149} {arXiv:hep-ph/0701149} \BibitemShut
  {NoStop}%
\end{thebibliography}%

\end{document}